\documentclass[12pt]{amsart}
\usepackage{amsmath}
\usepackage{amssymb}
\usepackage{tabularx}
\usepackage{enumerate}
\usepackage[dvipdfm]{graphicx}
\usepackage{texdraw}
\usepackage{color}
\usepackage{lineno}
\usepackage{tikz}
\usetikzlibrary {shapes.geometric}

\usepackage{mathrsfs}
\usepackage{amsfonts,amssymb,amsmath}
\usepackage{epsfig}
\usepackage{subfigure}
\usepackage{hyperref}
\usepackage{float}
\makeatletter
\@addtoreset{equation}{section}

\makeatother
\newtheorem{theorem}{Theorem}[section]

\newtheorem{corollary}{Corollary}[section]
\newtheorem{proposition}{Proposition}[section]

\begin{document}
\title[Bistable waves and propagation success on hexagonal lattice]
{Bistable waves and propagation success on hexagonal lattice$^\S$}
\thanks{$\S$ J. Fang is supported in part by the NSF of China (Nos. 12171119 and 12371164) and the Fundamental Research Funds for the Central Universities of China. Y. Li is supported in part by the NSF of China (No. 12301624) and China Postdoctoral Science Foundation grant (No. 2023M740934). Y. Lou is supported by The Hong Kong Research Grant Council (No. 15307823).}
\author[J. Fang, Y. Li, Y. Lou, J. Wang]{Jian Fang$^\dag$, Yifei Li$^\ddag$, Yijun Lou$\sharp$, Jian Wang$^\ddag$}
\thanks{$\dag$  Institute for Advanced Studies in Mathematics and School of Mathematics, Harbin Institute of Technology, Harbin 150001, China.}
\thanks{$\ddag$ School of Mathematics, Harbin Institute of Technology, Harbin 150001, China.}
\thanks{$\sharp$ Department of Applied Mathematics, The Hong Kong Polytechnic University, Hong Kong SAR, China.}
\thanks{E-mails: {\sf jfang@hit.edu.cn} (J. Fang), {\sf yifeili@hit.edu.cn} (Y. Li),  {\sf yijun.lou@polyu.edu.hk} (Y. Lou), {\sf jwang815@163.com} (J. Wang).}
\date{\today}

\begin{abstract}
Inspired by propagation phenomena occurring on media with hexagonal structures, we propose an idealized system of bistable reaction-diffusion equations posed on a hexagonal lattice, and investigate the existence and speed of traveling wave solutions. We establish a sharp criterion for propagation success (nonzero wave speed) and failure (zero wave speed), where the criterion is determined by an angle-dependent periodic threshold function and the ratio of two parameters in the nonlinearity. Moreover, the sign of the wave speeds is governed by the angle and the parameter ratio. Our results imply that the hexagonal structure and bistability significantly affect the propagation dynamics.
\end{abstract}

\subjclass[2010]{34A33, 35C07}
\keywords{hexagonal lattice, bistable equation, traveling wave, propagation success}
\maketitle

%%%%%%%%%%%%%%%%%%%%%%%%%%%%%%%%%%%%%%%%%%%%%%%%%%%%%%%%%%%%%%%%%%%%%%%%%%%%%%%%%%%%%%%%%%%%%%%%%%%%%%%%%%%%%%%%%%%%%

\section{Introduction and main results}\label{sec1}

Many spatial models have been developed to study how spatial structures and nonlinear interactions influence propagation dynamics and determine whether a wave will succeed or fail to propagate. For example, Bates and collaborators \cite{BC,BFRW} studied one-dimensional non-local bistable reaction-diffusion models arising from crystal phase transitions, and identified the criteria for propagation success and failure. The phenomenon of propagation failure (pinning) has also been described by one-dimensional reaction-diffusion equations with coupling terms in the context of nonlinear optics \cite{AAC, CER}. In addition to identifying the thresholds for wave propagation and extinction, the direction of propagation is another interesting issue. Motivated by crystal growth, Cahn et al. \cite{CMV} proposed a bistable reaction-diffusion equation posed on a square lattice, where new phenomena regarding the propagation direction were found. This is a remarkably distinct feature compared with the one-dimensional model. Besides crystal growth, Keener \cite{K} introduced a spatially discrete bistable reaction-diffusion equation for modeling signal transmission between excitable cells, in which the phenomenon of propagation failure was also identified. Tabi et al. \cite{TMM} used a differential-difference nonlinear equation to model long-distance propagation of intercellular $\mathrm{Ca^{2+}}$. 

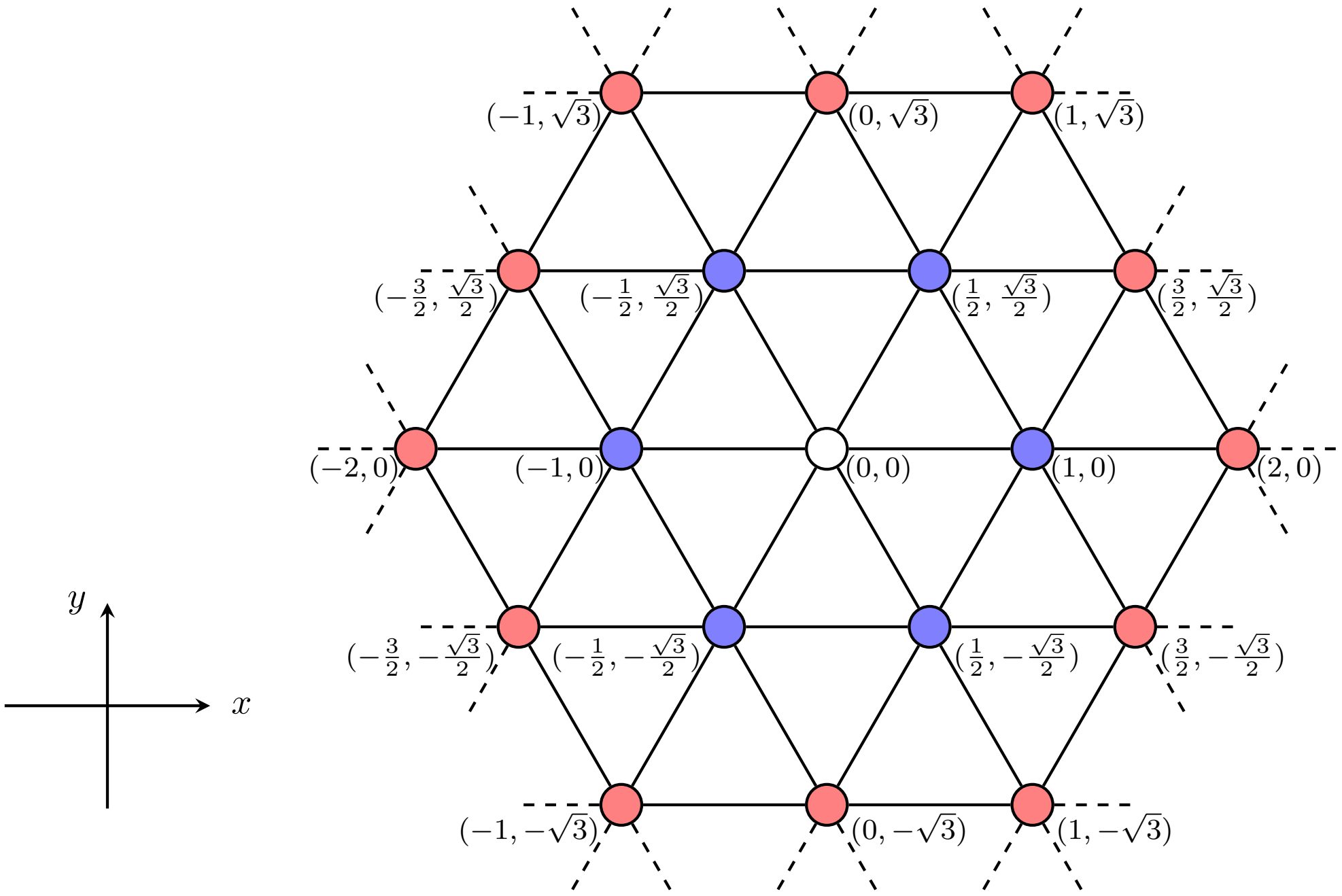
\begin{figure}[h]
	\centering
	\begin{tikzpicture}[>=stealth,thick]
		\node (s00) at (0,0) [circle, draw=black]{};
		\node (s10) at (2,0) [circle, draw, fill=blue!50]{};
		\node (s20) at (4,0) [circle, draw, fill=red!50]{};
		\draw [dashed,] (s20)--(4.5,-0.866); \draw [dashed,] (s20)--(4.5,0.866); \draw [dashed,] (s20)--(5,0);
		\node (s-10) at (-2,0) [circle, draw, fill=blue!50]{};
		\node (s-20) at (-4,0) [circle, draw, fill=red!50]{};
		\draw [dashed,] (s-20)--(-4.5,-0.866); \draw [dashed,] (s-20)--(-4.5,0.866); \draw [dashed,] (s-20)--(-5,0);
		\node (s-11) at (-3,1.732) [circle, draw, fill=red!50]{};
		\draw [dashed,](s-11)--(-3.5,2.598); \draw [dashed,](s-11)--(-4,1.732);
		\node (s01) at (-1,1.732) [circle, draw, fill=blue!50]{};
		\node (s11) at (1,1.732) [circle, draw, fill=blue!50]{};
		\node (s21) at (3,1.732) [circle, draw, fill=red!50]{};
		\draw [dashed,](s21)--(3.5,2.598); \draw [dashed,](s21)--(4,1.732);
		\node (s02) at (-2,3.464) [circle, draw, fill=red!50]{};
		\draw [dashed,](s02)--(-1.5,4.33); \draw [dashed,](s02)--(-2.5,4.33); \draw [dashed,](s02)--(-3,3.464);
		\node (s12) at (0,3.464) [circle, draw, fill=red!50]{};
		\draw [dashed,] (s12)--(-0.5,4.33); \draw [dashed,] (s12)--(0.5,4.33); 
		\node (s22) at (2,3.464) [circle, draw, fill=red!50]{};
		\draw [dashed,](s22)--(1.5,4.33); \draw [dashed,](s22)--(2.5,4.33); \draw [dashed,](s22)--(3,3.464);
		\node (s-2-1) at (-3,-1.732) [circle, draw, fill=red!50]{};
		\draw [dashed,](s-2-1)--(-3.5,-2.598); \draw [dashed,](s-2-1)--(-4,-1.732);
		\node (s-1-1) at (-1,-1.732) [circle, draw, fill=blue!50]{};
		\node (s0-1) at (1,-1.732) [circle, draw, fill=blue!50]{};
		\node (s1-1) at (3,-1.732) [circle, draw, fill=red!50]{};
		\draw [dashed,](s1-1)--(3.5,-2.598); \draw [dashed,](s1-1)--(4,-1.732);
		\node (s-2-2) at (-2,-3.464) [circle, draw, fill=red!50]{};
		\draw [dashed,](s-2-2)--(-1.5,-4.33); \draw [dashed,](s-2-2)--(-2.5,-4.33); \draw [dashed,](s-2-2)--(-3,-3.464);
		\node (s-1-2) at (0,-3.464) [circle, draw, fill=red!50]{};
		\draw [dashed,] (s-1-2)--(-0.5,-4.33); \draw [dashed,] (s-1-2)--(0.5,-4.33); 
		\node (s0-2) at (2,-3.464) [circle, draw, fill=red!50]{};
		\draw [dashed,](s0-2)--(1.5,-4.33); \draw [dashed,](s0-2)--(2.5,-4.33); \draw [dashed,](s0-2)--(3,-3.464);
		\draw [-] (s02)--(s12)node[]{};
		\draw [-] (s12)--(s22)node[]{};
		\draw [-] (s-11)--(s01)node[]{};
		\draw [-] (s01)--(s11)node[]{};
		\draw [-] (s11)--(s21)node[]{};
		\draw [-] (s-20)--(s-10)node[]{};
		\draw [-] (s-10)--(s00)node[]{};
		\draw [-] (s00)--(s10)node[]{};
		\draw [-] (s10)--(s20)node[]{};
		\draw [-] (s-2-1)--(s-1-1)node[]{};
		\draw [-] (s-1-1)--(s0-1)node[]{};
		\draw [-] (s0-1)--(s1-1)node[]{};
		\draw [-] (s-2-2)--(s-1-2)node[]{};
		\draw [-] (s-1-2)--(s0-2)node[]{};
		\draw [-] (s02)--(s-11)node[]{};
		\draw [-] (s-11)--(s-20)node[]{};
		\draw [-] (s12)--(s01)node[]{};
		\draw [-] (s01)--(s-10)node[]{};
		\draw [-] (s-10)--(s-2-1)node[]{};
		\draw [-] (s22)--(s11)node[]{};
		\draw [-] (s11)--(s00)node[]{};
		\draw [-] (s00)--(s-1-1)node[]{};
		\draw [-] (s-1-1)--(s-2-2)node[]{};
		\draw [-] (s21)--(s10)node[]{};
		\draw [-] (s10)--(s0-1)node[]{};
		\draw [-] (s0-1)--(s-1-2)node[]{};
		\draw [-] (s20)--(s1-1)node[]{};
		\draw [-] (s1-1)--(s0-2)node[]{};
		\draw [-] (s22)--(s21)node[]{};
		\draw [-] (s21)--(s20)node[]{};
		\draw [-] (s12)--(s11)node[]{};
		\draw [-] (s11)--(s10)node[]{};
		\draw [-] (s10)--(s1-1)node[]{};
		\draw [-] (s02)--(s01)node[]{};
		\draw [-] (s01)--(s00)node[]{};
		\draw [-] (s00)--(s0-1)node[]{};
		\draw [-] (s0-1)--(s0-2)node[]{};
		\draw [-] (s-11)--(s-10)node[]{};
		\draw [-] (s-10)--(s-1-1)node[]{};
		\draw [-] (s-1-1)--(s-1-2)node[]{};
		\draw [-] (s-20)--(s-2-1)node[]{};
		\draw [-] (s-2-1)--(s-2-2)node[]{};
		\node at (0.5,-0.2) []{\tiny$(0,0)$};
		\node at (2.5,-0.2) []{\tiny$(1,0)$};
		\node at (4.5,-0.2) []{\tiny$(2,0)$};
		\node at (-2.6,-0.2) []{\tiny$(-1,0)$};
		\node at (-4.6,-0.2) []{\tiny$(-2,0)$};
		\node at (-3.8,1.5) []{\tiny$(-\frac{3}{2},\frac{\sqrt{3}}{2})$};
		\node at (-1.8,1.5) []{\tiny$(-\frac{1}{2},\frac{\sqrt{3}}{2})$};
		\node at (1.7,1.5) []{\tiny$(\frac{1}{2},\frac{\sqrt{3}}{2})$};
		\node at (3.7,1.5) []{\tiny$(\frac{3}{2},\frac{\sqrt{3}}{2})$};
		\node at (-2.75,3.25) []{\tiny$(-1,\sqrt{3})$};
		\node at (0.65,3.25) []{\tiny$(0,\sqrt{3})$};
		\node at (2.65,3.25) []{\tiny$(1,\sqrt{3})$};
		\node at (-3.95,-2) []{\tiny$(-\frac{3}{2},-\frac{\sqrt{3}}{2})$};
		\node at (-1.95,-2) []{\tiny$(-\frac{1}{2},-\frac{\sqrt{3}}{2})$};
		\node at (1.85,-2) []{\tiny$(\frac{1}{2},-\frac{\sqrt{3}}{2})$};
		\node at (3.85,-2) []{\tiny$(\frac{3}{2},-\frac{\sqrt{3}}{2})$};	
		\node at (-2.9,-3.7) []{\tiny$(-1,-\sqrt{3})$};
		\node at (0.8,-3.7) []{\tiny$(0,-\sqrt{3})$};
		\node at (2.8,-3.7) []{\tiny$(1,-\sqrt{3})$};	
		\draw [->](-7,-3.5)--(-7,-1.5)node[left=0.05cm]{\footnotesize$y$};
		\draw [->](-8,-2.5)--(-6,-2.5)node[right=0.05cm]{\footnotesize$x$};		
	\end{tikzpicture}
	\caption{A hexagonal lattice described by the coordinates $x$-$y$.}\label{fig1}	
\end{figure}

As a common two-dimensional architecture in nature, the hexagonal lattice provides another spatial structure that governs both the dispersal patterns of individuals and the collective dynamics of the population \cite{FLLW}. The occurrence of propagation phenomena in media with hexagonal structures has been frequently reported in physics and cell biology, such as the growth front of hexagonal FeTe crystals \cite{KYZZ} and intercellular signaling in the entorhinal cortex \cite{HFM}. To characterize the mechanisms governing propagation success and failure, as well as the propagation direction in hexagonal architectures, we consider a bistable reaction-diffusion equation defined on a hexagonal lattice. The hexagonal lattice consists of nodes and edges connecting the nearest adjacent nodes, as shown in Figure \ref{fig1}, and the set of nodes is defined by
\begin{equation*}
	\mathbb{H}:=\bigcup_{i,j\in\mathbb{Z}}\left((i,\sqrt{3}j)\cup(\frac{1}{2}+i,\frac{\sqrt{3}}{2}+\sqrt{3}j)\right).
\end{equation*}
The resulting reaction-diffusion model on this discrete spatial structure is given by
\begin{equation}
	\label{eq1.1}
	u'(x,y,t)=\frac{1}{6}\Delta_h[u](x,y,t)+f(u(x,y,t)), \ (x,y)\in \mathbb{H},\ t>0,
\end{equation}
where $\Delta_h:\mathbb{R}\rightarrow\mathbb{R}$ is defined as 
\begin{equation*}
	%\label{eq1.2}
	\begin{aligned}
		&\Delta_h[u](x,y,\cdot):=u(x+1,y,\cdot)+u(x-1,y,\cdot)+u(x+\frac{1}{2},y+\frac{\sqrt{3}}{2},\cdot)\\
		&+u(x-\frac{1}{2},y-\frac{\sqrt{3}}{2},\cdot)+u(x-\frac{1}{2},y+\frac{\sqrt{3}}{2},\cdot)+u(x+\frac{1}{2},y-\frac{\sqrt{3}}{2},\cdot)-6u(x,y,\cdot),
	\end{aligned}				
\end{equation*}
and $f:\mathbb{R}\rightarrow\mathbb{R}$ is a piecewise linear function defined as
\begin{equation}
	\label{eq1.3}		
	f(u)=\left\{\begin{aligned}
		&-u,&&u\in[0,\alpha], \\
		&\beta-u, &&u\in(\alpha,\beta],
	\end{aligned}\right.
\end{equation}
with $0<\alpha<\beta$. Such a piecewise linear function was previously used for modeling phase transitions \cite{AC, KT}. The hexagonal lattice, equipped with the associated discrete diffusion $\Delta_h$, was first introduced in a recent work \cite{FLLW}, where the authors considered LDEs with a Fisher-KPP type nonlinearity. 

Spatial discretization of reaction-diffusion equations leads lattice differential equations (LDEs), which represent infinite systems of ordinary differential equations indexed over discrete lattice sites \cite{CMV2}. Propagation dynamics for LDEs, especially traveling waves, have been widely investigated. When the lattice is one-dimensional line structure, we refer to \cite{CG2, CGW, CMS, HS, Z1, Z2} for the existence, uniqueness and stability of traveling waves of monostable and bistable LDEs, \cite{K, MZ} for propagation and its failure in LDEs, \cite{FZ, LZ} for traveling waves of LDEs described by monotone semiflows, and \cite{CGH} for traveling waves of the epidemic model. When the lattice is two-dimensional square structure, Cheng et al. \cite{CLW} and Wu et al. \cite{WWR} studied delayed LDEs and showed that the spreading speed coincides with the minimal wave speed. Guo and Wu \cite{GW3} and Shen \cite{Sh} proved the existence of traveling waves of space-periodic and time-periodic LDEs, respectively. Hupkes and Van Vleck \cite{HV} established the existence of traveling waves of bistable LDEs with periodic and negative diffusion coefficients. In 1998, Cahn et al. \cite{CMV} proposed the following LDE on a square lattice:
\begin{equation}
	\label{eq11.1}	
	\begin{aligned}
		u'(i,j,t)=&u(i+1,j,t)+u(i-1,j,t)+u(i,j+1,t)+u(i,j-1,t)-4u(i,j,t)\\
		&+f(u(i,j,t)),\ (i,j)\in\mathbb{Z}^2,\ t>0,
	\end{aligned}					
\end{equation}
where $f$ is a piecewise function satisfying \eqref{eq1.3}. They revealed a complex relationship between the unique bistable wave speed and the propagation angle on $\mathbb{Z}^2$. Such a relationship is not seen in one-dimensional lattice. Later, Elmer and Van Vleck \cite{EV2} generalized some of the results in \cite{CMV} to $\mathbb{Z}^n$, $n\in\mathbb{N}^+$, from both theoretical and numerical perspectives. Compared to $\mathbb{Z}^n$, less effort has been made for the propagation dynamics of LDEs on more complicated discrete spatial structures. We refer to \cite{BF, LM} for the dynamics of diffusion equations on tree graphs. We further note that propagation failure in bistable LDEs has also been studied by localized solutions in \cite{CER}.

We now define the traveling wave solutions for \eqref{eq1.1}, which is the primary focus of this study. Let $\theta \in [0,2\pi)$ be the propagation angle relative to the lattice orientation, and define the unit direction vector $\mathrm{\bf{n}}=(\varsigma,\sigma):=(\cos\theta,\sin\theta)$. A traveling wave solution propagating in the direction $\mathrm{\bf{n}}$ is an ansatz of the form:
\begin{equation}	
	\label{eq1.40}		
	u(x,y,t)=U((x,y)\cdot(\varsigma,\sigma)+ct)=U(\xi), \ \  (x,y)\in\mathbb{H},
\end{equation}
where $c\in\mathbb{R}$ represents the wave speed and $U$ is the wave profile. Physically, the variable $\xi$ defines a co-moving frame. The dot product $(x,y)\cdot(\varsigma,\sigma)$ projects the two-dimensional lattice coordinates onto the direction of propagation, allowing the spatial evolution to be analyzed as a function of the single phase variable $\xi$. Substituting \eqref{eq1.40} into \eqref{eq1.1} yields the following equation
\begin{equation}
	\label{eq1.4}
	cU'(\xi)=\frac{1}{6}\sum_{k=1}^{3}\left(U(\xi+\delta_k)+U(\xi-\delta_k)-2U(\xi)\right)+f(U(\xi)), \ \xi\in\mathbb{R},
\end{equation}
where 
\begin{equation}
	\label{eq1.5}
	\delta_1:=\varsigma,\ \delta_2:=\frac{1}{2}\varsigma+\frac{\sqrt{3}}{2}\sigma, \  \delta_3:=\frac{1}{2}\varsigma-\frac{\sqrt{3}}{2}\sigma.
\end{equation}
We look for traveling wave solutions satisfying the following conditions:
\begin{equation}
	\label{c1.8}
	U(-\infty)=0, \ U(\infty)=\beta,\ U'(\xi)>0, \ U(0)=\alpha, \ c\neq0.
\end{equation}
Here we call such $U$ bistable waves, which are monotonic fronts connecting two equilibria $0$ and $\beta$. From \eqref{c1.8}, it follows that
\begin{equation*}
	U(\xi)<\alpha \ \text{for} \ \xi<0, \ \text{and} \ U(\xi)>\alpha \ \text{for}\ \xi>0.
\end{equation*}
Note that $f$ is discontinuous at $U=\alpha$. Let $H$ be the following Heaviside function: 
\begin{equation}
	\label{eq:h}
	H(v)=\left\{\begin{aligned}
		&0, &&v\leq0 ,\\
		&\beta, && v>0.
	\end{aligned}\right.
\end{equation}
Define a linear operator $\mathcal{D}_h:C^1(\mathbb{R})\rightarrow C^1(\mathbb{R})$ with
\begin{equation*}	
	\mathcal{D}_h[U](\xi)=\sum_{k=1}^{3}\left(U(\xi+\delta_k)+U(\xi-\delta_k)-2U(\xi)\right),
\end{equation*}
where $\delta_1, \delta_2$ and $\delta_3$ are defined in \eqref{eq1.5}. Combining \eqref{eq1.4} and \eqref{c1.8}, we can conclude that $U(\xi)$ satisfies
\begin{equation}
	\label{eq1.6}
	\left\{\begin{aligned}
		&cU'(\xi)=\frac{1}{6}\mathcal{D}_h[U](\xi)+H(U(\xi)-\alpha)-U(\xi), && \xi\in\mathbb{R},\\
		&U(-\infty)=0, \ U(0)=\alpha,\  U(\infty)=\beta. &&
	\end{aligned}\right.
\end{equation}
By looking for the solution of \eqref{eq1.1} satisfying the specific form \eqref{eq1.40} defined by traveling wave coordinate, we have transformed the two-dimensional reaction-diffusion equation \eqref{eq1.1} on discrete space into the one-dimensional equation \eqref{eq1.6} on continuous space incorporating the propagation angle on the hexagonal lattice. In this work, we focus on identifying the range of angles and parameters that admit solutions of \eqref{eq1.6}, which are the traveling wave solutions of \eqref{eq1.1}.

Before presenting the main results of this paper, we define the following function
\begin{equation}
	\label{eq1.51}	
	\begin{aligned}
		G(s,\theta):=&1-\frac{1}{3}\Big[\cos(s\cos\theta)+\cos\Big(\frac{1}{2}s\cos\theta-\frac{\sqrt{3}}{2}s\sin\theta\Big)\\
		&+\cos\Big(\frac{1}{2}s\cos\theta+\frac{\sqrt{3}}{2}s\sin\theta\Big)-3\Big], \ s\in\mathbb{R},\ \theta\in[0,2\pi),
	\end{aligned}	
\end{equation}   
which is introduced as an angle-dependent auxiliary function. It plays a fundamental role in characterizing the wave speed, the critical thresholds for propagation and pinning, and the underlying directionality of the wave motion.

\begin{theorem}
	\label{th1.1}
	Let $G(s,\theta)$ be defined in \eqref{eq1.51}. The limit
	\begin{equation*}
		%\label{eq2.00}	
		\gamma(\theta)=\lim_{T\rightarrow\infty}\frac{1}{2T}\int_{0}^{T}\frac{1}{G(s,\theta)}\mathrm{d}s, \ \theta\in[0,2\pi)
	\end{equation*}
	exists. Further, the following statements hold:  
	\begin{itemize}
		\item[$\mathrm{(i)}$] $\gamma(\theta)=\gamma(\theta+\frac{\pi}{3})\in(\frac{1}{6},\frac{1}{2})$ for $\theta\in[0,2\pi)$.
		\item[$\mathrm{(ii)}$] $\gamma(\theta)$ is continuous at each $\theta$ where $\tan\theta\in\mathbb{R}\backslash\{\sqrt{3}r:r\in\mathbb{Q}\}$, and is discontinuous at each $\theta$ where $\tan\theta\in\{\sqrt{3}r:r\in\mathbb{Q}\}\cup\{\pm\infty\}$.
	\end{itemize}	 
\end{theorem}

\begin{theorem}
	\label{th1.2}
	Let $\gamma(\theta)$ be defined in Theorem \ref{th1.1}. If $\tan\theta\in\mathbb{R}\backslash\{\sqrt{3}r:r\in\mathbb{Q}\}$, then \eqref{eq1.1} admits no traveling waves. If $\tan\theta\in\{\sqrt{3}r:r\in\mathbb{Q}\}\cup\{\pm\infty\}$, then the following statements hold: 
	\begin{itemize}
		\item[$\mathrm{(i)}$] If $|\frac{\alpha}{\beta}-\frac{1}{2}|>\gamma(\theta)$, then \eqref{eq1.1} admits a bistable wave $U(x\varsigma+y\sigma+ct)$, where $c\neq0$ is a unique solution of
		\begin{equation}
			\label{eq2.1}	
			\frac{\alpha}{\beta}-\frac{1}{2}=-\frac{1}{\pi}\int_{0}^{\infty}\frac{c}{G(s,\theta)^2+c^2s^2}\mathrm{d}s.
		\end{equation} 
		\item[$\mathrm{(ii)}$] If $|\frac{\alpha}{\beta}-\frac{1}{2}|\leq\gamma(\theta)$, then \eqref{eq1.1} admits no bistable waves with $c\neq0$.
	\end{itemize}			
\end{theorem}
We refer to the existence of the bistable wave with the nonzero wave speed as propagation success, and the absence of such bistable wave as propagation failure. This theorem provides a sharp classification of the scenarios leading to propagation success and propagation failure, as well as a characterization of the wave speed with respect to propagation angle. 

The remainder of this paper is organized as follows. In Section~\ref{sec2}, an explicit traveling wave solution is first obtained using the Fourier transform method developed in \cite{CMV}. Theorem~\ref{th1.1} is then proved by properties of almost periodic functions, and Theorem~\ref{th1.2} follows from a combination of the explicit solution and Theorem~\ref{th1.1}. In Section~\ref{sec3}, we present numerical investigations to further illustrate the relationship between the wave speed and propagation angle. Section~\ref{sec4} serves as the conclusion of this paper.
 
\section{Proofs of the theorems}\label{sec2}

We apply the methods in \cite{CMV} for \eqref{eq11.1} to our equation \eqref{eq1.1}. Assume that $U$ is a solution of \eqref{eq1.6} for some $c\ne0$. Denote $V(\xi):=U(-\xi)$. Then
\begin{equation}
	\label{c2.1}
	V(-\infty)=\beta, \quad V(\infty)=0,\quad V(0)=\alpha.
\end{equation}
The definition of  the Heaviside function $H$ in \eqref{eq:h} gives $H(V(\xi)-\alpha)=0$ for $\xi>0$, which, combining with the first equation of \eqref{eq1.6}, implies that 
\begin{equation}
	\label{eq2.2}	
	-cV'(\xi)=\frac{1}{6}\mathcal{D}_h[V](\xi)-V(\xi), \ \xi>0.
\end{equation}
Introduce the Laplace transform as
\begin{equation}
	\label{eq2.3}	
	\widehat{V}(s)=\int_{0}^{\infty}\mathrm{e}^{-s\xi}V(\xi)\mathrm{d}\xi.
\end{equation}
Using \eqref{c2.1} and \eqref{eq2.3}, we obtain from \eqref{eq2.2} that
\begin{equation*}
	\label{eq2.4}	
	cV(0)=\int_{0}^{\infty}\frac{1}{6}\mathcal{D}_h[V](\xi)\mathrm{e}^{-s\xi}\mathrm{d}\xi-(1-cs)\widehat{V}(s).
\end{equation*}
By \eqref{eq2.3} again, it follows that
\begin{equation*}
	\label{eq2.5}	
	\begin{aligned}
		&\int_{0}^{\infty}\frac{1}{6}\mathcal{D}_h[V](\xi)\mathrm{e}^{-s\xi}\mathrm{d}\xi\\
		=&\frac{1}{6}\int_{0}^{\infty}\left[\sum_{k=1}^{3}V(\xi+\delta_k)\mathrm{e}^{-s\xi}+V(\xi-\delta_k)\mathrm{e}^{-s\xi}-2V(\xi)\mathrm{e}^{-s\xi}\right]\mathrm{d}\xi\\
		=&\frac{1}{6}\sum_{k=1}^{3}\bigg[\int_{0}^{\infty}V(\xi)\mathrm{e}^{-s\xi}\mathrm{e}^{s\delta_k}\mathrm{d}\xi-\int_{0}^{\delta_k}V(\xi)\mathrm{e}^{-s\xi}\mathrm{e}^{s\delta_k}\mathrm{d}\xi+\int_{0}^{\infty}V(\xi)\mathrm{e}^{-s\xi}\mathrm{e}^{-s\delta_k}\mathrm{d}\xi\\
		&+\int_{-\delta_k}^{0}V(\xi)\mathrm{e}^{-s\xi}\mathrm{e}^{-s\delta_k}\mathrm{d}\xi-2\int_{0}^{\infty}\mathrm{e}^{-s\xi}V(\xi)\mathrm{d}\xi\bigg]\\
		=&\frac{1}{6}\sum_{k=1}^{3}\left[\left(\mathrm{e}^{s\delta_k}+\mathrm{e}^{-s\delta_k}-2\right)\widehat{V}(s)-\mathrm{e}^{s\delta_k}\int_{0}^{\delta_k}\mathrm{e}^{-s\xi}V(\xi)\mathrm{d}\xi-\mathrm{e}^{-s\delta_k}\int_{-\delta_k}^{0}\mathrm{e}^{-s\xi}V(\xi)\mathrm{d}\xi\right].
	\end{aligned}
\end{equation*}
Combining the above two identities, we further obtain 
\begin{equation*}
	\widehat{V}(s)=\frac{-cV(0)-\frac{1}{6}\sum_{k=1}^{3}\left(\mathrm{e}^{s\delta_k}\int_{0}^{\delta_k}\mathrm{e}^{-s\xi}V(\xi)\mathrm{d}\xi-\mathrm{e}^{-s\delta_k}\int_{-\delta_k}^{0}\mathrm{e}^{-s\xi}V(\xi)\mathrm{d}\xi\right)}
	{1-cs-\frac{1}{6}\sum_{k=1}^{3}\left(\mathrm{e}^{s\delta_k}+\mathrm{e}^{-s\delta_k}-2\right)}.
\end{equation*}
Using the same arguments as in \cite[Lemma 4.1]{CMV}, we can infer that, for $c\neq0$, there exist $\varepsilon_0>0$ and $K>0$ such that
\begin{equation}
	\label{eq2.6}
	|U(\xi)|\leq K\mathrm{e}^{\varepsilon_0\xi}, \ \xi\leq0.
\end{equation}

Let $\mathrm{i}$ be the imaginary unit and consider the Fourier transform
\begin{equation}
	\label{eq2.7}	
	\widehat{U}_\varepsilon(s)=\int_{-\infty}^{\infty}\mathrm{e}^{-\mathrm{i}s\xi}U_\varepsilon(\xi)\mathrm{d}\xi, \quad U_\varepsilon(\xi)=\mathrm{e}^{-\varepsilon\xi}U(\xi),
\end{equation}
where $0<\varepsilon<\varepsilon_0$, and $\varepsilon_0$ satisfies \eqref{eq2.6}. Then \eqref{eq2.6} implies that $U_\varepsilon(\xi)\rightarrow0$ as $\xi\rightarrow-\infty$. By \eqref{eq2.7}, it follows that $U_\varepsilon(\xi)\rightarrow0$ as $\xi\rightarrow\infty$. Multiplying both sides of the first equation of \eqref{eq1.6} by $\mathrm{e}^{-\varepsilon\xi}$, we obtain
\begin{equation}
	\label{eq2.8}
	\begin{aligned}	
		cU'_\varepsilon(\xi)=&\frac{1}{6}\sum_{k=1}^{3}\left[\mathrm{e}^{\varepsilon\delta_k}U_\varepsilon(\xi+\delta_k)+\mathrm{e}^{-\varepsilon\delta_k}U_\varepsilon(\xi-\delta_k)-2U_\varepsilon(\xi)\right]-(1+c\varepsilon)U_\varepsilon(\xi)\\
		&+\mathrm{e}^{-\varepsilon\xi}H(U(\xi)-\alpha).
	\end{aligned}	
\end{equation}
By the definition of  function $H$ in \eqref{eq:h}, it follows that
\begin{equation*}
	\int_{-\infty}^{\infty}\mathrm{e}^{-(\mathrm{i}s+\varepsilon)\xi}H(U(\xi)-\alpha)\mathrm{d}\xi =\frac{\beta}{\mathrm{i}s+\varepsilon}.
\end{equation*}
Multiplying both sides of \eqref{eq2.8} by $\mathrm{e}^{-\mathrm{i}s\xi}$, and then integrating the resulting expression with respect to $\xi$ from $-\infty$ to $\infty$, we have
\begin{equation}
	\label{eq2.9}
	\begin{aligned}
		\widehat{U}_\varepsilon(s)&=\frac{\beta}{(\mathrm{i}s+\varepsilon)\left[1+(\mathrm{i}s+\varepsilon)c-\frac{1}{6}\sum_{k=1}^{3}(\mathrm{e}^{(\mathrm{i}s+\varepsilon)\delta_k}+\mathrm{e}^{-(\mathrm{i}s+\varepsilon)\delta_k}-2)\right]}\\
		&=\frac{\beta}{(\mathrm{i}s+\varepsilon)W(s-\mathrm{i}\varepsilon,\theta)},
	\end{aligned}
\end{equation}
where 
\begin{equation}
	\label{eq2.10}
	W(s,\theta):=\mathrm{i}cs+1-\frac{1}{3}\Big[\cos\varsigma s+\cos\Big(\frac{1}{2}\varsigma-\frac{\sqrt{3}}{2}\sigma\Big)s+\cos\Big(\frac{1}{2}\varsigma+\frac{\sqrt{3}}{2}\sigma\Big)s-3\Big].
\end{equation}
From the Fourier inversion theorem, we infer that $U(\xi)$ can be written as an absolutely convergent integral
\begin{equation*}
	U(\xi)=\mathrm{e}^{\varepsilon\xi}U_\varepsilon(\xi)=\frac{1}{2\pi}\int_{-\infty}^{\infty}\mathrm{e}^{(\mathrm{i}s+\varepsilon)\xi}\widehat{U}_\varepsilon(s)\mathrm{d}s,
\end{equation*}
which, combining with \eqref{eq2.9}, yields 
\begin{equation}
	\label{eq2.11}		
	U(\xi)=\frac{\beta}{2\pi\mathrm{i}}\int_{-\infty-\mathrm{i}\varepsilon}^{\infty-\mathrm{i}\varepsilon}\frac{\mathrm{e}^{\mathrm{i}s\xi}}{sW(s,\theta)}\mathrm{d}s.
\end{equation}
The limit of the above integration when $\varepsilon\rightarrow0$ represents the contour integral along a horizontal line. For any fixed $\theta$ and small $\varepsilon$, the integrand $\frac{\mathrm{e}^{\mathrm{i}s\xi}}{sW(s,\theta)}$ is holomorphic in the horizontal strip $-\varepsilon\leq \mathrm{Im}(s)\leq0$ except for a simple pole $s=0$.  Note that $\mathrm{e}^{\mathrm{i}s\xi}$ is bounded in $-\varepsilon\leq \mathrm{Im}(s)\leq0$ for any fixed $\xi\geq0$, and $\frac{1}{sW(s,\theta)}=O(|s|^{-2})$ for all $\theta\in[0,2\pi)$ as $|\mathrm{Re}(s)|\rightarrow\infty$ in $-\varepsilon\leq \mathrm{Im}(s)\leq0$. To evaluate \eqref{eq2.11} via the residue theorem, we introduce a contour consisting of the intervals $L_\varepsilon:=(-\infty,-\varepsilon]\cup[\varepsilon,\infty)$ along the real axis and a semi-circular arc $J_\varepsilon:=\{\varepsilon\mathrm{e}^{\mathrm{i}\iota}|\iota\in[-\pi,0]\}$ in the complex plane. Then, for any fixed $\theta$, we shift the contour of integration in \eqref{eq2.11} to get
\begin{equation}
	\label{eq2.12}		
	U(\xi)=\frac{\beta}{2\pi\mathrm{i}}\left(\int_{L_\varepsilon}+\int_{J_\varepsilon}\right)\frac{\mathrm{e}^{\mathrm{i}s\xi}}{sW(s,\theta)}\mathrm{d}s
\end{equation}
with a contour direction that increases $\mathrm{Re}(s)$.

We replace $L_\varepsilon$ with $(-\infty,-\varepsilon]$ and $[\varepsilon,\infty)$ in the real axis to get
\begin{equation}
	\label{eq2.13}	
	\begin{aligned}	
		\frac{\beta}{2\pi\mathrm{i}}\int_{L_\varepsilon}\frac{\mathrm{e}^{\mathrm{i}s\xi}}{sW(s,\theta)}\mathrm{d}s&=\frac{\beta}{2\pi\mathrm{i}}\left(\int_{-\infty}^{-\varepsilon}+\int_{\varepsilon}^{\infty}\right)\frac{\mathrm{e}^{\mathrm{i}s\xi}}{sW(s,\theta)}\mathrm{d}s\\
		&=\frac{\beta}{2\pi\mathrm{i}}\int_{\varepsilon}^{\infty}\left[\frac{\mathrm{e}^{-\mathrm{i}s\xi}}{-sW(-s,\theta)}+\frac{\mathrm{e}^{\mathrm{i}s\xi}}{sW(s,\theta)}\right]\mathrm{d}s\\
		&=-\frac{\beta c}{\pi}\int_{\varepsilon}^{\infty}\frac{\cos(s\xi)}{c^2s^2+G(s,\theta)^2}\mathrm{d}s+\frac{\beta}{\pi}\int_{\varepsilon}^{\infty}\frac{G(s,\theta)\sin(s\xi)}{s(c^2s^2+G(s,\theta)^2)}\mathrm{d}s,
	\end{aligned}	
\end{equation}
where $G(s,\theta)$, $\theta\in[0,2\pi)$, is given by \eqref{eq1.51}. From \eqref{eq2.10}, we see that $G(s,\theta)$ can be regarded as the real part of $W(s,\theta)$ for the real number $s$ and all $\theta\in[0,2\pi)$. Note that $J_\varepsilon$ is a half-circle and $s=0$ is a simple pole. For every $\theta\in[0,2\pi)$, the residue theorem implies 
\begin{equation}
	\label{eq2.14}	
	\frac{\beta}{2\pi\mathrm{i}}\int_{J_\varepsilon}\frac{\mathrm{e}^{\mathrm{i}s\xi}}{sW(s,\theta)}\mathrm{d}s=\frac{\beta}{2}\ \text{as}\ \varepsilon\rightarrow0.
\end{equation}
By substituting \eqref{eq2.13} and \eqref{eq2.14} into \eqref{eq2.12}, and sending $\varepsilon\rightarrow0$, we can conclude that
\begin{equation}
	\label{eq21.1}	
	U(\xi)=\frac{\beta}{2}+\frac{\beta}{\pi}\int_{0}^{\infty}\frac{G(s,\theta)\sin(s\xi)}{s(G(s,\theta)^2+c^2s^2)}\mathrm{d}s-\frac{\beta c}{\pi}\int_{0}^{\infty}\frac{\cos(s\xi)}{G(s,\theta)^2+c^2s^2}\mathrm{d}s.
\end{equation}

\begin{proposition}
	\label{pr1}	
	If \eqref{eq21.1} is a solution of \eqref{eq1.6}, then $U'>0$ on $\mathbb{R}$. Moreover, $U'(\xi)$ is discontinuous at $\xi=0$.
\end{proposition}
\begin{proof}
We start with the following claim.

\textit{Claim}: The complex function $W(s,\theta)$ defined in \eqref{eq2.10} admits a simple root $s=-b_0\mathrm{i}$ with $b_0>0$, and $\mathrm{i}\frac{\partial W(s,\theta)}{\partial s}\big{|}_{s=-b_0\mathrm{i}}>0$ for any fixed $\theta\in[0,2\pi)$. Furthermore, if $s=a-b\mathrm{i}$, $b\geq0$, is the other root of $W(s,\theta)$ for any fixed $\theta\in[0,2\pi)$, then $b>b_0$.

\textit{The proof of Claim}: Let $b\geq0$. Then we substitute $s=-b\mathrm{i}$ into $W(s,\theta)$ to get
\begin{equation}
	\label{eq2.18}	
	\begin{aligned}	
		W(-b\mathrm{i},\theta)&=bc+1-\frac{1}{3}\left[\sum_{k=1}^{3}\frac{1}{2}(\mathrm{e}^{b\delta_k}+\mathrm{e}^{-b\delta_k})-3\right]\\
		&=bc+1-\frac{1}{3}\left[\sum_{k=1}^{3}\cosh(b\delta_k)-3\right]\\
		&:=R(b,\theta),\ \theta\in[0,2\pi).
	\end{aligned}	
\end{equation}
It is easy to get
\begin{equation*}
	\frac{\partial R(b,\theta)}{\partial b}=c-\frac{1}{3}\sum_{k=1}^{3}\delta_k\sinh(b\delta_k), \ \text{and}\ \frac{\partial^2 R(b,\theta)}{\partial b^2}=-\frac{1}{3}\sum_{k=1}^{3}\delta^2_k\cosh(b\delta_k)<0.
\end{equation*}
Note that $W(0,\theta)=R(0,\theta)=1$ and $W(-b\mathrm{i},\theta)<0$ for large $b$ and all $\theta\in[0,2\pi)$. Then we infer that there exists a unique $b_0$ such that $R(b_0,\theta)=0$ for all $\theta\in[0,2\pi)$. Thus, for every $\theta\in[0,2\pi)$, we have 
\begin{equation}
	\label{eq2.20}	
	W(-b\mathrm{i},\theta)>0 \ \text{on}\ [0,b_0), \ W(-b_0\mathrm{i},\theta)=0,\ W(-b\mathrm{i},\theta)<0 \ \text{on}\ (b_0,\infty).
\end{equation}
From $W(-b_0\mathrm{i},\theta)=0$, it follows that $-b_0c=2-\frac{1}{6}\sum_{k=1}^{3}(\mathrm{e}^{b_0\delta_k}+\mathrm{e}^{-b_0\delta_k})$. For all $\theta\in[0,2\pi)$, we obtain 
\begin{equation*}
	\begin{aligned}	
		b_0\mathrm{i}\frac{\partial W(s,\theta)}{\partial s}\bigg{|}_{s=-b_0\mathrm{i}}&=b_0\mathrm{i}\left[c\mathrm{i}+\frac{1}{3}\sum_{k=1}^{3}\delta_k\sin(-b_0\delta_k\mathrm{i})\right]\\
		&=1+1-\frac{1}{6}\sum_{k=1}^{3}(\mathrm{e}^{b_0\delta_k}+\mathrm{e}^{-b_0\delta_k})+\frac{1}{6}\sum_{k=1}^{3}b_0\delta_k(\mathrm{e}^{b_0\delta_k}-\mathrm{e}^{-b_0\delta_k})\\
		&:=1+Q(\delta_k), 
	\end{aligned}	
\end{equation*}
where $k=1$, $2$, $3$. Since $\delta_k(\mathrm{e}^{b_0\delta_k}-\mathrm{e}^{-b_0\delta_k})\geq0$ for $b_0>0$ and $\delta_k\in[-1,1]$, $k=1$, $2$, $3$, we have
\begin{equation*}
	Q''(\delta_k)=\frac{1}{6}\sum_{k=1}^{3}\left(b_0^2(\mathrm{e}^{b_0\delta_k}+\mathrm{e}^{-b_0\delta_k})+b_0^3\delta_k(\mathrm{e}^{b_0\delta_k}-\mathrm{e}^{-b_0\delta_k})\right)>0.
\end{equation*}
It is easily verified that $Q(\delta_k)\geq Q(0)=0$ for $\delta_k\in[-1,1]$, $k=1$, $2$, $3$. Thus we get $\mathrm{i}\frac{\partial W(s,\theta)}{\partial s}\big{|}_{s=-b_0\mathrm{i}}>0$.

If $s:=a-b\mathrm{i}$ is another root of $W(s,\theta)$ for any fixed $\theta\in[0,2\pi)$, $a\in\mathbb{R}$, $b\geq0$, then it follows that
\begin{equation}
	\label{eq2.21}	
	\begin{aligned}	
		0&=W(a-b\mathrm{i},\theta)\\
		&=\mathrm{i}ac+bc+1-\frac{1}{3}\left[\sum_{k=1}^{3}\cos(a\delta_k-b\delta_k\mathrm{i})-3\right]\\
		&=bc+1-\frac{1}{3}\left[\sum_{k=1}^{3}\cos(a\delta_k)\cosh(b\delta_k)-3\right]+\left[ac-\frac{1}{3}\sum_{k=1}^{3}\sin(a\delta_k)\sinh(b\delta_k)\right]\mathrm{i}.
	\end{aligned}	
\end{equation}
Comparing the real part of $W(a-b\mathrm{i},\theta)$ in \eqref{eq2.21} and \eqref{eq2.18}, we have
\begin{equation}
	\label{eq2.22}	
	0=\mathrm{Re}(W(a-b\mathrm{i},\theta))\geq W(-b\mathrm{i},\theta) \ \text{for all} \ \theta\in[0,2\pi).
\end{equation}
We obtain from \eqref{eq2.20} that $b\geq b_0$. Due to \eqref{eq2.21}, it follows that $\mathrm{Im}(W(a-b\mathrm{i},\theta))=0$, that is,
\begin{equation}
	\label{eq2.23}	
	ac=\frac{1}{3}\sum_{k=1}^{3}\sin(a\delta_k)\sinh(b\delta_k),
\end{equation}
which implies that, for all roots $s$ in the strip $b_0\leq b\leq b_0+1$, $a$ in $\mathrm{Re}(s)$ is bounded, and $W(s,\theta)$ possesses a finite number of roots in the strip. Thus, we infer that the only root on the horizontal line $\mathrm{Im}(s)=-b_0$ of the strip is $s=-b_0\mathrm{i}$. 

Note that $b\geq b_0$. To prove $b>b_0$, we assume, for contradiction, that $b=b_0$. Taking $b=b_0$ in \eqref{eq2.22} and combining with \eqref{eq2.21} and \eqref{eq2.18},  we obtain $\cos(a\delta_k)=1$, $k=1$, $2$, $3$, which implies that $\sin(a\delta_k)=0$, $k=1$, $2$, $3$, in \eqref{eq2.21}. From \eqref{eq2.23}, it follows that $a=0$, that is, $s=-b_0\mathrm{i}$ is another root of $W(s,\theta)$ for  fixed $\theta\in[0,2\pi)$, which leads to a contradiction. This means that $b>b_0$. Hence the \textit{Claim} holds.

Based on this \textit{Claim}, we use techniques similar to those on proof in \cite[Corollary 4.5 and Theorem 4.6]{CMV} to obtain that $U'(\xi)>0$ on $\mathbb{R}$.

Next, we will show that $U'(\xi)$ is discontinuous at $\xi=0$. It is evident that the left and right limits of the solution $U(\xi)$ at $0$ exist. Taking $\xi\rightarrow0^-$ and $\xi\rightarrow0^+$ in the right side of the first equation of \eqref{eq1.6}, respectively, we get
\begin{equation*}
	\lim_{\xi\rightarrow0^-}cU'(\xi)=\frac{1}{6}\mathcal{D}_h[U](0)-U(0), \  \lim_{\xi\rightarrow0^+}cU'(\xi)=\frac{1}{6}\mathcal{D}_h[U](0)+\beta-U(0),
\end{equation*}
which implies that $\lim_{\xi\rightarrow0^+}U'(\xi)-\lim_{\xi\rightarrow0^-}U'(\xi)=\frac{\beta}{c}$, $c\neq0$. Therefore, $U'(\xi)$ is discontinuous at $\xi=0$. This completes the proof. 
\end{proof}

\begin{figure}[ht]
	\centering
	\vspace{-10pt}
	{\includegraphics[width=2.8in,height=2.15in,clip]{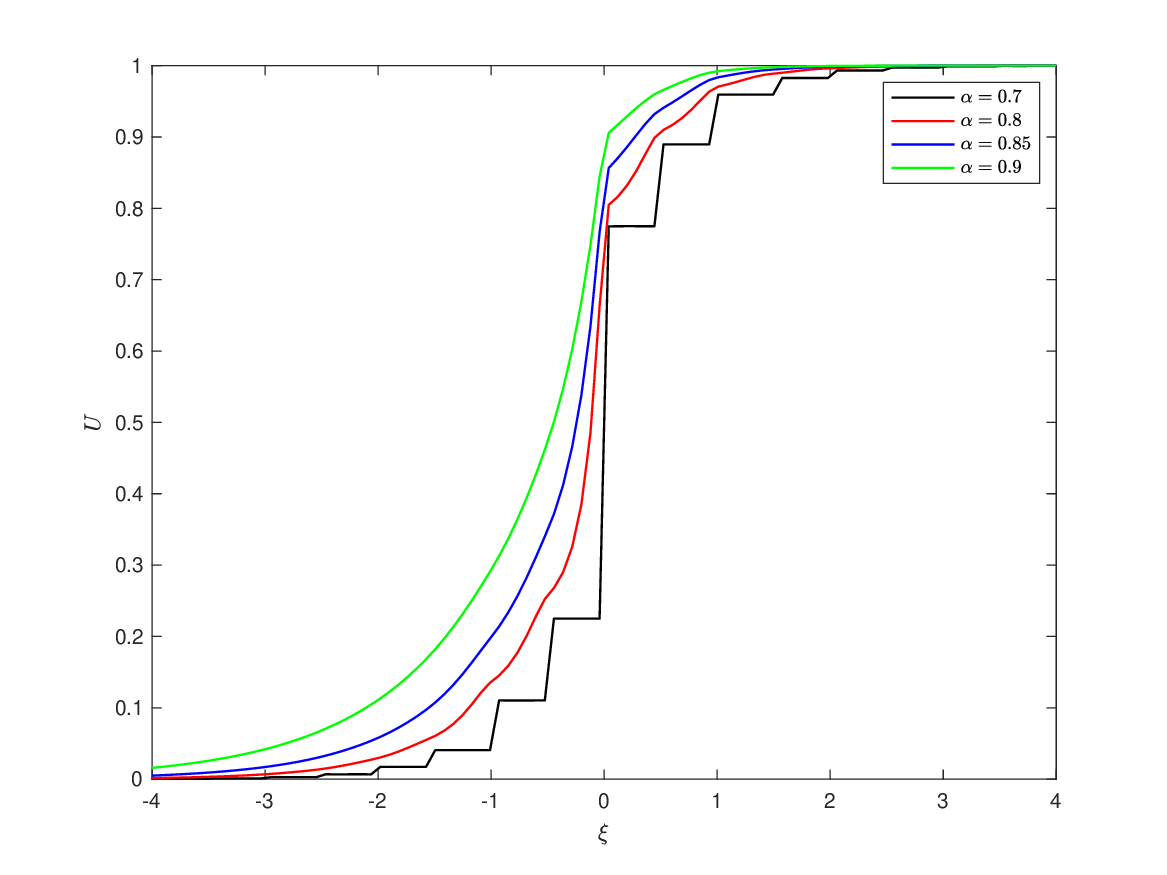} }	
	\vspace{-10pt}
	\caption{The relationship between the wave profile $U$ and $\xi$ in \eqref{eq21.1} for $\beta=1$, $\theta=0$ and four distinct $\alpha$.}\label{fig2}	
\end{figure}

\noindent \textbf{Numerical illustration of Proposition \ref{pr1}.} We plot the profile of $U(\xi)$ with $\beta=1$ while varying $\alpha$ values in  Figure \ref{fig2}.  It shows that $U(\xi)$ is strictly increasing with $\xi$, and has a `kink' at $\xi=0$ for $\alpha=0.8$, $0.85$, $0.9$. The `kink' implies that $U'(0^-)\neq U'(0^+)$. Moreover, $U(\xi)$ shows a step-like behavior for $\alpha=0.7$.

\noindent \textbf{Proof of Theorem \ref{th1.1}}. We first show that $\lim_{T\rightarrow\infty}\frac{1}{2T}\int_0^T\frac{1}{G(s,\theta)}\mathrm{d}s$ exists for all $\theta\in[0,2\pi)$, where $G(s,\theta)$ is given by \eqref{eq1.51}. From the definition of $G(s,\theta)$, we infer that $G(s,\theta)\in[1,3]$ for  $s\in\mathbb{R}$ and $\theta\in[0,2\pi)$.

The notations in \eqref{eq1.5} imply 
\begin{equation}
	\label{eq3.1}
	\delta_1=\cos\theta, \ \delta_2=\frac{1}{2}\cos\theta+\frac{\sqrt{3}}{2}\sin\theta, \ \delta_3=\frac{1}{2}\cos\theta-\frac{\sqrt{3}}{2}\sin\theta,\ \theta\in[0,2\pi).
\end{equation}
Then $G(s,\theta)$ defined by \eqref{eq1.51} can be rewritten as 
\begin{equation}
	\label{eq1.52}
	G(s,\theta)=1-\frac{1}{3}\left[\cos(s\delta_1)+\cos(s\delta_2)+\cos(s\delta_3)-3\right], \ s\in\mathbb{R},\ \theta\in[0,2\pi).
\end{equation}
When $\delta_k=0$, $k=1$, $2$, $3$, it is easily verified that $G$ is a periodic function of $s\in\mathbb{R}$. When $\delta_k\neq0$, $k=1$, $2$, $3$, then, for any $\epsilon>0$ and $T_\epsilon>0$, there exists $n_k\in\mathbb{Z}$, $k=1$, $2$, $3$, such that $|\delta_kT_\epsilon-2\pi n_k|\leq\epsilon$. Moreover, there exists $\epsilon_0\in[-2\epsilon,0]$ such that $\delta_kT_\epsilon=2\pi n_k+\epsilon+\epsilon_0$, $k=1$, $2$, $3$. Then, for fixed $\theta$, we have
\begin{equation*}
	\begin{aligned}	
		|\cos(\delta_k(s+T_\epsilon))-\cos(\delta_ks)|&=	|\cos(\delta_ks+\epsilon+\epsilon_0+2\pi n_k)-\cos(\delta_ks)|\\
		&=|\cos(\delta_ks+\epsilon+\epsilon_0)-\cos(\delta_ks)|\\
		&=\left|-2\sin(\delta_ks+\frac{\epsilon+\epsilon_0}{2})\sin\frac{\epsilon+\epsilon_0}{2}\right|\\
		&\leq\epsilon+\epsilon_0\leq\epsilon, \ s\in\mathbb{R},
	\end{aligned}	
\end{equation*}	
which implies that $\cos(\delta_ks)$ is almost periodic in $s\in\mathbb{R}$ for all $\theta\in[0,2\pi)$, where $k=1$, $2$, $3$. Expression \eqref{eq1.52} and $G(s,\theta)\in[1,3]$ imply that $G(s,\theta)$ is almost periodic in $s\in\mathbb{R}$ for all $\theta\in[0,2\pi)$. From \cite[Theorem 3.4]{C}, we can conclude that $\gamma(\theta)=$ $\lim_{T\rightarrow\infty}\frac{1}{2T}\int_0^T\frac{1}{G(s,\theta)}\mathrm{d}s$ exists for all $\theta\in[0,2\pi)$. 

\noindent\textbf{Proof of (i)}. 
Combining the definition of $G$ with $\gamma(\theta)$, we calculate that $\gamma(\theta)=\gamma(\theta+\frac{\pi}{3})\in(\frac{1}{6},\frac{1}{2})$ for all $\theta\in[0,2\pi)$.  

\noindent\textbf{Proof of (ii)}.
Applying $\cos z=\frac{1}{2}(\mathrm{e}^{\mathrm{i}z}+\mathrm{e}^{-\mathrm{i}z})$, $z\in\mathbb{R}$, to \eqref{eq1.52}, we rewrite the integrand $\frac{1}{G(s,\theta)}$ in $\gamma(\theta)$ as the Fourier series:
\begin{equation}
	\label{eq3.2}	
	\begin{aligned}		
		\frac{1}{G(s,\theta)}&=\frac{1}{2}\frac{1}{1-\frac{1}{12}(\mathrm{e}^{-\mathrm{i}s\delta_1}+\mathrm{e}^{\mathrm{i}s\delta_1}+\mathrm{e}^{-\mathrm{i}s\delta_2}+\mathrm{e}^{\mathrm{i}s\delta_2}+\mathrm{e}^{-\mathrm{i}s\delta_3}+\mathrm{e}^{\mathrm{i}s\delta_3})}\\
		&=\frac{1}{2}\sum_{k=0}^{\infty}\left(\frac{1}{12}\right)^k(\mathrm{e}^{-\mathrm{i}s\delta_1}+\mathrm{e}^{\mathrm{i}s\delta_1}+\mathrm{e}^{-\mathrm{i}s\delta_2}+\mathrm{e}^{\mathrm{i}s\delta_2}+\mathrm{e}^{-\mathrm{i}s\delta_3}+\mathrm{e}^{\mathrm{i}s\delta_3})^k\\
		&=\sum_{l=-\infty}^{\infty}\sum_{m=-\infty}^{\infty}\sum_{n=-\infty}^{\infty}\Phi_{l,m,n}\mathrm{e}^{\mathrm{i}s(l\delta_1+m\delta_2+n\delta_3)},
	\end{aligned}
\end{equation}
where $\Phi_{l,m,n}>0$ is the Fourier coefficients, $l,m,n\in\mathbb{Z}$. It is easily verified that the sums in \eqref{eq3.2} are absolutely convergent. This, together with $\gamma(\theta)$, yields
\begin{equation}
	\label{eq3.3}	
	\begin{aligned}	
		\gamma(\theta)&=\lim_{T\rightarrow\infty}\frac{1}{2T}\int_{0}^{T}\sum_{l=-\infty}^{\infty}\sum_{m=-\infty}^{\infty}\sum_{n=-\infty}^{\infty}\Phi_{l,m,n}\mathrm{e}^{\mathrm{i}s(l\delta_1+m\delta_2+n\delta_3)}\mathrm{d}s\\
		&=\frac{1}{2}\sum_{l\delta_1+m\delta_2+n\delta_3=0}\Phi_{l,m,n}.
	\end{aligned}
\end{equation}
Subsequently, we use \eqref{eq3.3} to investigate the continuity and discontinuity of $\gamma(\theta)$ with respect to $\theta\in[0,2\pi)$ in two cases.

\textit{Case 1}. When $\tan\theta=\pm\infty$, we use \eqref{eq3.1} to get $(\delta_1, \delta_2, \delta_3)=(0, \pm\frac{\sqrt{3}}{2},\mp\frac{\sqrt{3}}{2})$. By $l\delta_1+m\delta_2+n\delta_3=0$, it follows that $\frac{\sqrt{3}}{2}m=\frac{\sqrt{3}}{2}n$, $m,n\in\mathbb{Z}$. 

When $\tan\theta=\sqrt{3}r$, $r\in\mathbb{Q}$, we obtain from \eqref{eq3.1} that $\delta_1=\frac{1}{\sqrt{1+3r^2}}$, $\delta_2=\frac{1+3r}{2\sqrt{1+3r^2}}$ and $\delta_3=\frac{1-3r}{2\sqrt{1+3r^2}}$, $r\in\mathbb{Q}$. Due to $l\delta_1+m\delta_2+n\delta_3=0$, it follows that
\begin{equation}
	\label{eq3.4}	
	l=\frac{3r}{2}(n-m)-\frac{1}{2}(n+m), \ l,m,n\in\mathbb{Z}.
\end{equation}
It is evident that \eqref{eq3.4} holds for $l=-m$, $m=n\in\mathbb{Z}$. Assume that $\theta$ satisfies 
\begin{equation*}
	\tan\theta\in\{\sqrt{3}r\}\cup\{\pm\infty\}, \ r\in\mathbb{Q}.
\end{equation*}
Then we infer that $l\delta_1+m\delta_2+n\delta_3=0$ holds only for $l=-m=-n$, $l,m,n\in\mathbb{Z}$. Expression \eqref{eq3.3} implies that 
\begin{equation*}		
	\gamma(\theta)=\frac{1}{2}\sum_{l\delta_1+m\delta_2+n\delta_3=0}\Phi_{l,m,n}, \ l=-m=-n, \ l,m,n\in\mathbb{Z}. 
\end{equation*}
This means that the values of $l$, $m$ and $n$ satisfying $l\delta_1+m\delta_2+n\delta_3=0$ depend on $\theta$. Therefore, $\gamma(\theta)$ can be regarded as a discontinuous function of $\theta\in[0,2\pi)$ when $\tan\theta\in\{\sqrt{3}r:r\in\mathbb{Q}\}\cup\{\pm\infty\}$. 

\textit{Case 2}. When $\tan\theta\in(\mathbb{R}\backslash\mathbb{Q})\backslash\sqrt{3}r$, $r\in\mathbb{Q}$, we can conclude from \eqref{eq3.1} that 
\begin{equation*}
	(\cos\theta,\sin\theta)\in\left\{(\mathbb{Q}, \mathbb{R}\backslash\mathbb{Q}),\  (\mathbb{R}\backslash\mathbb{Q}, \mathbb{R}\backslash\mathbb{Q}),\  (\mathbb{R}\backslash\mathbb{Q}, \mathbb{Q})\right\},
\end{equation*}
which implies that $\delta_1\neq\delta_2\neq\delta_3$, $\delta_2\neq-\delta_3$ and
\begin{equation}
	\label{eq3.7}	
	(\delta_1, \delta_2, \delta_3)\in\left\{(\mathbb{Q}, \mathbb{R}\backslash\mathbb{Q}, \mathbb{R}\backslash\mathbb{Q}), \ (\mathbb{R}\backslash\mathbb{Q}, \mathbb{R}\backslash\mathbb{Q}, \mathbb{R}\backslash\mathbb{Q})\right\}.
\end{equation} 

When $\tan\theta\in\mathbb{Q}\backslash0$, we can infer from \eqref{eq3.1} that 
\begin{equation*}
	(\cos\theta,\sin\theta)\in\left\{(\mathbb{Q}\backslash0, \mathbb{Q}\backslash0),\  (\mathbb{R}\backslash\mathbb{Q}, \mathbb{R}\backslash\mathbb{Q})\right\},
\end{equation*}
which implies that $\delta_1\neq\delta_2\neq\delta_3$, $\delta_2\neq-\delta_3$ and
\begin{equation}
	\label{eq3.8}	
	(\delta_1, \delta_2, \delta_3)\in\left\{(\mathbb{Q}\backslash0, \mathbb{R}\backslash\mathbb{Q}, \mathbb{R}\backslash\mathbb{Q}), \ (\mathbb{R}\backslash\mathbb{Q}, \mathbb{R}\backslash\mathbb{Q}, \mathbb{R}\backslash\mathbb{Q})\right\}.
\end{equation} 
Suppose that $\theta$ satisfies
\begin{equation*}			
	\tan\theta\in\{(\mathbb{R}\backslash\mathbb{Q})\backslash\sqrt{3}r\}\cup\{\mathbb{Q}\backslash0\}=\mathbb{R}\backslash\sqrt{3}r, \ r\in\mathbb{Q}.
\end{equation*}
Then \eqref{eq3.7} and \eqref{eq3.8} imply that $l\delta_1+m\delta_2+n\delta_3=0$ holds true only when $l=m=n=0$. Due to \eqref{eq3.3}, it follows that
$\gamma(\theta)=\frac{\Phi_{0,0,0}}{2}$. Since $\Phi_{0,0,0}$ is a constant, $\gamma(\theta)$ is continuous of $\theta\in[0,2\pi)$ when $\tan\theta\in\mathbb{R}\backslash\{\sqrt{3}r:r\in\mathbb{Q}\}$. 

From \textit{Case 1} and \textit{Case 2}, we can conclude that statement (ii) holds. The proof of Theorem \ref{th1.1} is completed.

\begin{figure}[ht]
	\centering
	\vspace{-10pt}
	{\includegraphics[width=2.6in,height=2.3in,clip]{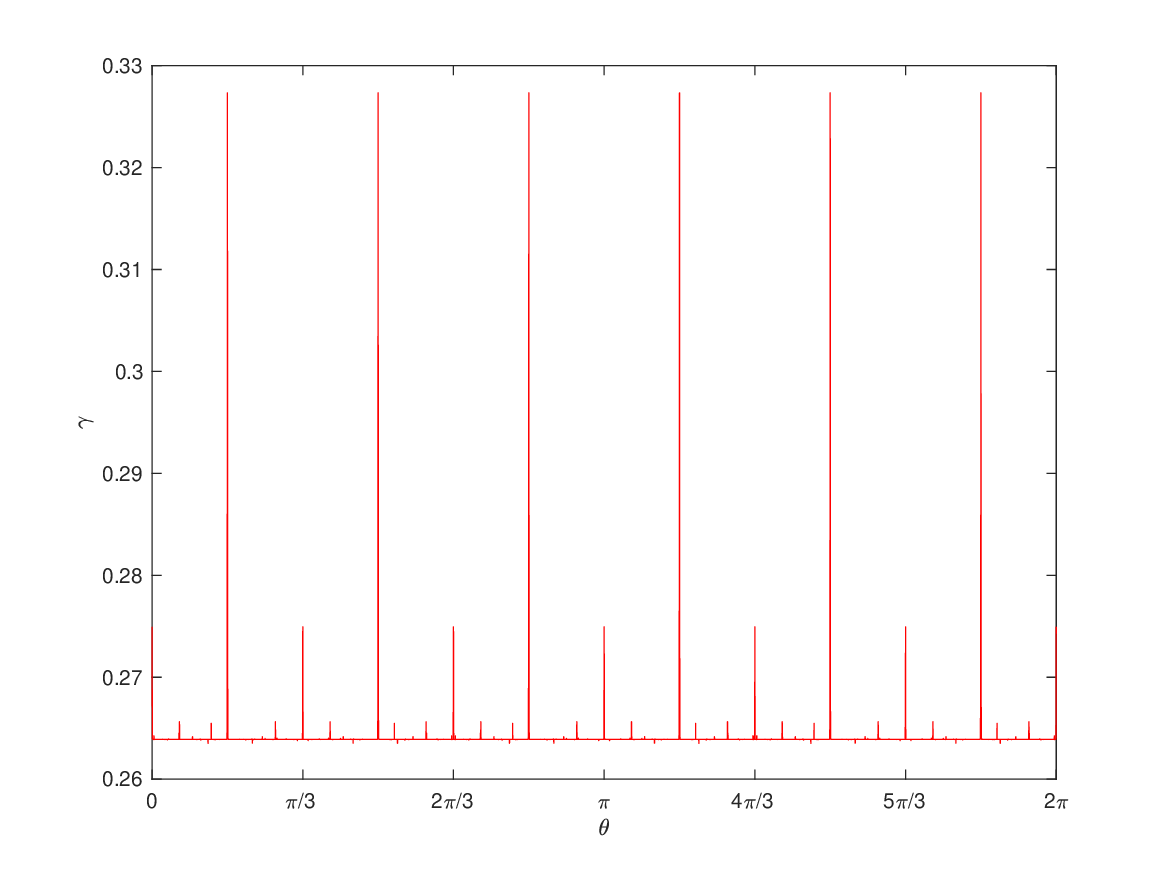} }	
	\hspace{-20pt}
	{\includegraphics[width=2.6in,height=2.3in,clip]{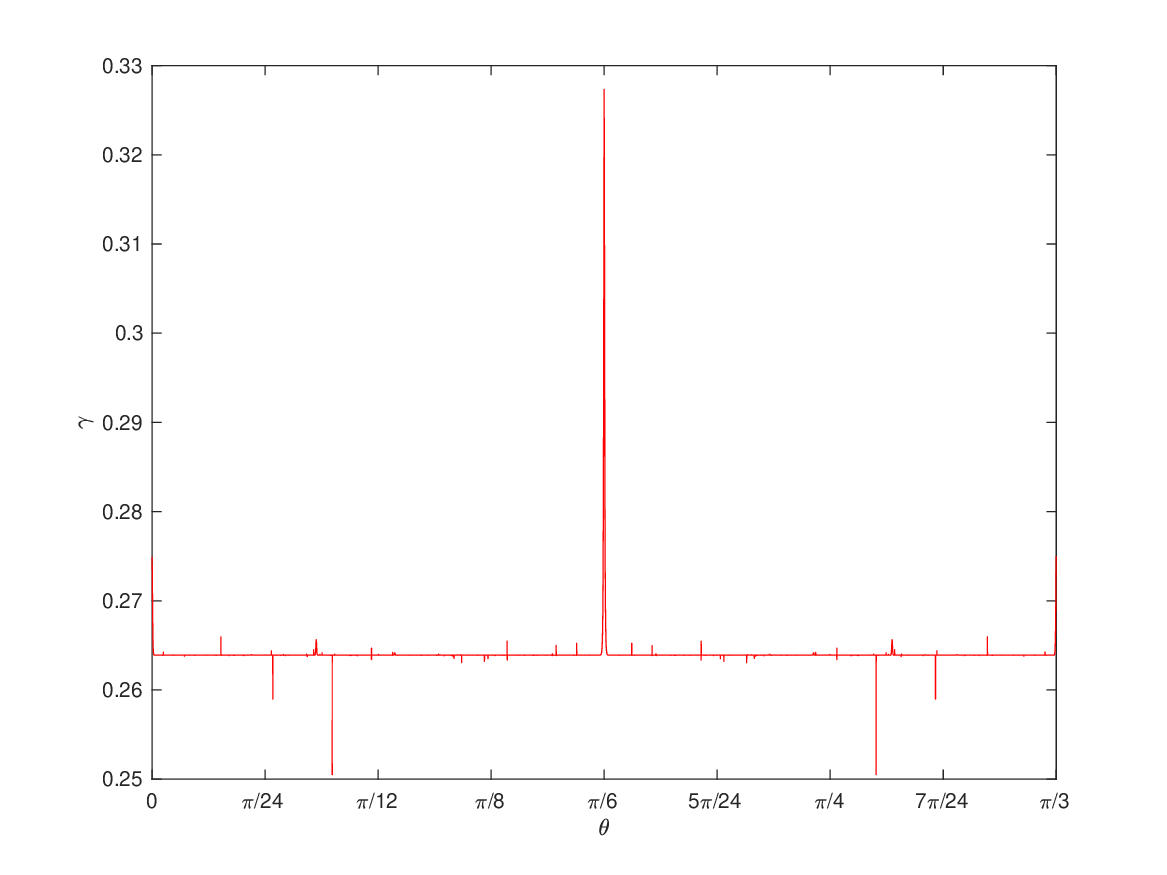} }	
	\vspace{-10pt}
	\caption{The relationship between $\gamma(\theta)$ and $\theta$ in Theorem \ref{th1.1}.}\label{fig4}	
\end{figure}

\noindent \textbf{Numerical illustration of Theorem \ref{th1.1}.}  To illustrate the profile of the function $\gamma(\theta)$ with respect to $\theta$, we use $\gamma(\theta) = \lim_{c \rightarrow 0^-} \Gamma(c, \theta)$, where $\Gamma(c, \theta)$ is given in equation~\eqref{eq2.250} and will be established in Theorem~\ref{th1.2} below. For numerical purposes, we choose $\gamma(\theta) = \Gamma(-0.0006, \theta)$ and consider $400000$ equidistant points on the intervals $[0, 2\pi]$ and $[0, \frac{\pi}{3}]$, respectively. The relationship between $\gamma(\theta)$ and $\theta$ is depicted in Figure~\ref{fig4}. The left panel shows that $\frac{\pi}{3}$ is the period of $\gamma(\theta)$ with respect to $\theta$. The plot of $\gamma(\theta)$ on $[0,\frac{\pi}{3}]$ is given in the right panel, where $\gamma(\theta)$ has a lot of `jumps', which imply that $\gamma(\theta)$ is discontinuous at some $\theta$.

\noindent\textbf{Proof of Theorem \ref{th1.2}}. According to the geometry of hexagonal lattice in Figure \ref{fig1}, each node in $\mathbb{H}$ can be represented by $(\frac{p}{2},\frac{\sqrt{3}q}{2})$, where $p$ and $q$ are both odd or both even. For $u$ acting on $\mathbb{H}\times[0,\infty)$, the direction in which $u$ on the node $(\frac{p}{2},\frac{\sqrt{3}q}{2})$ moves toward its nearest neighbor is
\begin{equation*}
	\tan\theta=\sqrt{3}\frac{q}{p}:=\sqrt{3}r \ \text{or}\ \tan\theta=\pm\infty, \ r\in\mathbb{Q}.
\end{equation*}
That is, \eqref{eq1.1} has no bistable waves when $\tan\theta\in\mathbb{R}\backslash\{\sqrt{3}r:r\in\mathbb{Q}\}$.

When $\tan\theta\in\{\sqrt{3}r:r\in\mathbb{Q}\}\cup\{\pm\infty\}$, we find conditions to ensure that \eqref{eq21.1} is a bistable wave of \eqref{eq1.1}. Indeed, $U(0)=\alpha$ implies that 
\begin{equation}
	\label{eq22.1}	
	\frac{\alpha}{\beta}-\frac{1}{2}=-\frac{ c}{\pi}\int_{0}^{\infty}\frac{1}{G(s,\theta)^2+c^2s^2}\mathrm{d}s.
\end{equation}
To facilitate further analysis of \eqref{eq22.1}, we define 
\begin{equation}
	\label{eq2.250}	
	\Gamma(c,\theta):=-\frac{c}{\pi}\int_{0}^{\infty}\frac{1}{G(s,\theta)^2+c^2s^2}\mathrm{d}s, \ \theta\in[0,2\pi).
\end{equation}

Next, we prove statements \textbf{(i)} and \textbf{(ii)} by investigating the properties of $\Gamma(c,\theta)$ in the above expression. The proofs are divided into the following three steps.

\textit{Step 1}. We will prove $\frac{\partial\Gamma(c,\theta)}{\partial c}<0$ for $\theta\in[0,2\pi)$ and $c\neq0$. Note that $1\leq G(s,\theta)\leq3$ for all $s\in\mathbb{R}$ and $\theta\in[0,2\pi)$. When $c\in(0,1]$, there exists a positive constant $\epsilon$ such that $\epsilon\leq c$, and, for all $\theta\in[0,2\pi)$, we get
\begin{equation}
	\label{eq2.16}		
	|G(s,\theta)^2+c^2s^2|\geq1+c^2s^2\geq c^2+c^2s^2\geq\epsilon^2(1+s^2) \ \mathrm{for}\ s\geq0.
\end{equation}
When $c\in[1,\infty)$, for all $\theta\in[0,2\pi)$, we also get
\begin{equation}
	\label{eq2.17}		
	|G(s,\theta)^2+c^2s^2|\geq1+c^2s^2\geq 1+s^2 \ \mathrm{for}\ s\geq0.
\end{equation}
Thus, each of the integrands in \eqref{eq21.1} has an upper bound for all $\theta\in[0,2\pi)$, $c\neq0$ and $\xi\in\mathbb{R}$. Note that each of such integrands is differentiable with respect to $c\neq0$ and $\theta\in[0,2\pi)$. This implies that $U(\xi)$ and $\Gamma(c,\theta)$ are differentiable with respect to $c\neq0$ for all $\theta\in[0,2\pi)$, $\xi\in\mathbb{R}$ and the fixed $\beta$. Moreover, $\Gamma(c,\theta)$ with $c\neq0$ is also differentiable with respect to $\theta\in[0,2\pi)$.

Let $P(\xi):=\frac{\partial U(\xi)}{\partial c}$, $c\neq0$. By differentiating both sides of \eqref{eq21.1} with respect to $c$ under the integral sign, and combining \eqref{eq2.16} with \eqref{eq2.17}, we infer that $P(\xi)$ is bounded with respect to $c\neq0$ as $\xi\rightarrow\pm\infty$ and $\xi\neq0$. Note that $U(\xi)$ given by \eqref{eq21.1} is continuously differentiable with respect to $c\neq0$. We can conclude from \eqref{eq1.6} that $U'(\xi)$ is also continuously differentiable with respect to $c\neq0$. Thus it is easy to prove that $\frac{\partial^2U(\xi)}{\partial c\partial\xi}=\frac{\partial (U'(\xi))}{\partial c}$. By differentiating both sides of the first equation of \eqref{eq1.6} with respect to $c$, we obtain
\begin{equation*}
	U'(\xi)+c\frac{\partial (U'(\xi))}{\partial c}=\frac{1}{6}\sum_{k=1}^{3}\left(\frac{\partial U(\xi+\delta_k)}{\partial c}+\frac{\partial U(\xi-\delta_k)}{\partial c}-2\frac{\partial U(\xi)}{\partial c}\right)-\frac{\partial U(\xi)}{\partial c},
\end{equation*}
which implies
\begin{equation}
	\label{eq2.27}		
	U'(\xi)+cP'(\xi)=\frac{1}{6}\mathcal{D}_h[P](\xi)-P(\xi), \ \xi\neq0.
\end{equation}
Therefore we get $P'(\xi)=\frac{\partial (U'(\xi))}{\partial c}$. By \eqref{eq22.1} and \eqref{eq2.250}, it follows that $\frac{U(0)}{\beta}=\frac{1}{2}$+$\Gamma(c,\theta)$. This implies that $\frac{\partial \Gamma(c,\theta)}{\partial c}=P(0)$ for all $\theta\in[0,2\pi)$ and $c\neq0$.

Let $Q(\xi):=U'(-\xi)$, $\xi\in\mathbb{R}$. Then $Q(\xi)$ is discontinuous at $\xi=0$, and we can infer from \eqref{eq1.6} that $Q$ is continuously differentiable on $(-\infty,0)\cup(0,\infty)$ and $Q(\pm\infty)=0$. Note that $\lim_{\xi\rightarrow0^+}U'(\xi)-\lim_{\xi\rightarrow0^-}U'(\xi)=\frac{1}{c}$. Then we obtain 
\begin{equation}
	\label{eq2.28}		
	Q(0^-)-Q(0^+)=U'(0^+)-U'(0^-)=\frac{1}{c}, \ c\neq0.
\end{equation}
The right side of the first equation of \eqref{eq1.6} implies that $U'(\xi)$ is differentiable with respect to $\xi$. It then follows that
\begin{equation}
	\label{eq2.29}		
	-cQ'(\xi)=\frac{1}{6}\mathcal{D}_h[Q](\xi)-Q(\xi), \ \xi\neq0.
\end{equation}
Moreover, $Q(\xi)$ and $Q'(\xi)$ are integrable on $\mathbb{R}$. Multiplying both sides of \eqref{eq2.27} by $Q(\xi)$, and then integrating the resulting expression from $-\infty$ to $\infty$, we have
\begin{equation}
	\label{eq2.30}	
	\begin{aligned}
		&\int_{-\infty}^{\infty}(U'(\xi)+cP'(\xi))Q(\xi)\mathrm{d}\xi=\int_{-\infty}^{\infty}\left(\frac{1}{6}\mathcal{D}_h[P](\xi)-P(\xi)\right)Q(\xi)\mathrm{d}\xi\\
		=&\frac{1}{6}\sum_{k=1}^{3}\left[\int_{-\infty}^{\infty}P(\xi+\delta_k)Q(\xi)\mathrm{d}\xi+\int_{-\infty}^{\infty}P(\xi-\delta_k)Q(\xi)\mathrm{d}\xi-2\int_{-\infty}^{\infty}P(\xi)Q(\xi)\mathrm{d}\xi\right]\\
		&-\int_{-\infty}^{\infty}P(\xi)Q(\xi)\mathrm{d}\xi\\
		=&\frac{1}{6}\sum_{k=1}^{3}\left[\int_{-\infty}^{\infty}P(\xi)Q(\xi-\delta_k)\mathrm{d}\xi+\int_{-\infty}^{\infty}P(\xi)Q(\xi+\delta_k)\mathrm{d}\xi-2\int_{-\infty}^{\infty}P(\xi)Q(\xi)\mathrm{d}\xi\right]\\
		&-\int_{-\infty}^{\infty}P(\xi)Q(\xi)\mathrm{d}\xi\\
		=&\int_{-\infty}^{\infty}\left(\frac{1}{6}\mathcal{D}_h[Q](\xi)-Q(\xi)\right)P(\xi)\mathrm{d}\xi.
	\end{aligned}
\end{equation}
From the boundedness of $P(\pm\infty)$, $Q(\pm\infty)=0$, \eqref{eq2.28} and \eqref{eq2.29}, it follows that
\begin{equation}
	\label{eq2.31}	
	\begin{aligned}
		&c\int_{-\infty}^{\infty}P'(\xi)Q(\xi)\mathrm{d}\xi\\
		=&cP(\xi)Q(\xi)|_{-\infty}^0-c\int_{-\infty}^{0}P(\xi)Q'(\xi)\mathrm{d}\xi+cP(\xi)Q(\xi)|^{\infty}_0-c\int_{0}^{\infty}P(\xi)Q'(\xi)\mathrm{d}\xi\\
		=&cP(0)(Q(0^-)-Q(0^+))-c\int_{-\infty}^{\infty}P(\xi)Q'(\xi)\mathrm{d}\xi\\
		=&P(0)+\int_{-\infty}^{\infty}P(\xi)\left(\frac{1}{6}\mathcal{D}_h[Q](\xi)-Q(\xi)\right)\mathrm{d}\xi.
	\end{aligned}
\end{equation}
Since $\frac{\partial \Gamma(c,\theta)}{\partial c}=P(0)$, we obtain from \eqref{eq2.30} and \eqref{eq2.31} that
\begin{equation*}
	\frac{\partial \Gamma(c,\theta)}{\partial c}=-\int_{-\infty}^{\infty}U'(\xi)Q(\xi)\mathrm{d}\xi=-\int_{-\infty}^{\infty}U'(\xi)U'(-\xi)\mathrm{d}\xi.
\end{equation*}
Since $c$ is the wave speed of $U(\xi)$, $-c$ is the wave speed of $U(-\xi)$. By \eqref{eq21.1}, it follows that $U(\xi)=\beta-U(-\xi)$, which, together with $U'(\xi)>0$, implies that $\frac{\partial\Gamma(c,\theta)}{\partial c}<0$ for $\theta\in[0,2\pi)$ and $c\neq0$.

\textit{Step 2}. We will show that $|\Gamma(c,\theta)|<\frac{1}{2}$,   
\begin{equation}
	\label{eq2.32}	
	\lim_{c\rightarrow-\infty}\Gamma(c,\theta)=\frac{1}{2}\ \text{and} \ \lim_{c\rightarrow0^-}\Gamma(c,\theta)=\gamma(\theta) \ \text{for all}\ \theta\in[0,2\pi)\ \text{and} \ c\neq0.
\end{equation}
By the definition of $\Gamma(c,\theta)$, given by \eqref{eq2.250}, it follows that
\begin{equation*}
	|\Gamma(c,\theta)|=\frac{|c|}{\pi}\int_{0}^{\infty}\frac{1}{G(s,\theta)^2+c^2s^2}\mathrm{d}s<\frac{|c|}{\pi}\int_{0}^{\infty}\frac{1}{1+c^2s^2}\mathrm{d}s=\frac{1}{2}
\end{equation*}
for all $\theta\in[0,2\pi)$ and  $c\neq0$. 

Although the dependence of $\Gamma(-c,\theta)$ and $G(s,\theta)$ on $\theta$ has not yet been discussed, in the following proof of~\eqref{eq2.32}, we note that the analysis of $\Gamma(-c,\theta)$ and $G(s,\theta)$ is valid for all $\theta \in [0, 2\pi)$, We introduce the following notation for simplicity:
\begin{equation*}
	\Gamma(-c,\theta):=\Lambda(c), \ G(s,\theta):=G(s)\ \text{and} \ \gamma(\theta):=\gamma \ \text{for all}\ \theta\in[0,2\pi)\ \text{and} \ c\neq0.
\end{equation*}
Then it follows that
\begin{equation}
	\label{eq2.33}		
	\Lambda(c)=\frac{c}{\pi}\int_{0}^{\infty}\frac{1}{G(s)^2+c^2s^2}\mathrm{d}s=\frac{1}{\pi}\int_{0}^{\infty}\frac{1}{G(\frac{s}{c})^2+s^2}\mathrm{d}s,\ c\neq0.
\end{equation}
Note that $G(\frac{s}{c})\rightarrow G(0)=1$ as $c\rightarrow\infty$. Combining the Lebesgue dominated theorem, we infer that $\lim_{c\rightarrow-\infty}\Gamma(c,\theta)=\lim_{c\rightarrow\infty}\Lambda(c)=\frac{1}{2}$ for  $\theta\in[0,2\pi)$. 

To prove the second limit in \eqref{eq2.32}, we always assume $c>0$ in \eqref{eq2.33}. Note that $G(s)$ is an almost periodic function with respect to $s\in\mathbb{R}$. According to the property of almost periodic functions \cite{S}, for any $\varepsilon>0$, there exists $M_\varepsilon$ that satisfies $M_\varepsilon\rightarrow\infty$ as $\varepsilon\rightarrow0$ such that each interval of length $M_\varepsilon$ consists of an $\varepsilon$-translation $T_\varepsilon\in[M_\varepsilon,2M_\varepsilon]$, and
\begin{equation}
	\label{eq2.34}	
	|G(s)-G(s+T_\varepsilon)|\leq\varepsilon, \ s\in\mathbb{R}.
\end{equation} 
Before taking the limit $\varepsilon \rightarrow 0$, we first fix $\varepsilon$ and assume $\varepsilon \leq \frac{1}{4}$ in the following arguments. Define $T_\varepsilon$-periodic function $G_\varepsilon(s):\mathbb{R}\rightarrow\mathbb{R}$ as follows: 
\begin{equation*}
	G_\varepsilon(s)=G(s)\ \text{for} \ s\in[0,T_\varepsilon), \ G_\varepsilon(s+T_\varepsilon)=G_\varepsilon(s) \ \text{for} \ s\in\mathbb{R}.
\end{equation*}
When $s\in[-T_\varepsilon,0)$, we have $s+T_\varepsilon\in[0,T_\varepsilon)$, which, together with \eqref{eq2.34}, yields 
\begin{equation*}
	|G(s)-G_\varepsilon(s)|=|G(s)-G_\varepsilon(s+T_\varepsilon)|=|G(s)-G(s+T_\varepsilon)|\leq\varepsilon.
\end{equation*}
Thus we infer that 
\begin{equation}
	\label{eq2.35}	
	|G(s)-G_\varepsilon(s)|\leq\varepsilon, \  s\in[-T_\varepsilon,T_\varepsilon).
\end{equation}
For any fixed $n\in\mathbb{N}$, let $\rho_n\in[-(n+1)T_\varepsilon,-nT_\varepsilon]$ be an $\varepsilon$-translation number of $G(s)$. Then $T_\varepsilon$ is the length of the interval in which $\rho_n$ resides. This means that 
\begin{equation}
	\label{eq2.36}	
	|G(s)-G(s+\rho_n)|\leq\varepsilon, \ \rho_n\in[-(n+1)T_\varepsilon,-nT_\varepsilon],\ s\in\mathbb{R}.
\end{equation} 
For any $s\in[nT_\varepsilon,(n+1)T_\varepsilon)$, we have
\begin{equation*}
	-T_\varepsilon=nT_\varepsilon-(n+1)T_\varepsilon\leq s+\rho_n<(n+1)T_\varepsilon-nT_\varepsilon=T_\varepsilon,
\end{equation*}
which, combined with \eqref{eq2.35} and \eqref{eq2.36}, implies 
\begin{equation}
	\label{eq2.37}	
	|G(s)-G_\varepsilon(s+\rho_n)|\leq|G(s)-G(s+\rho_n)|+|G(s+\rho_n)-G_\varepsilon(s+\rho_n)|\leq\varepsilon+\varepsilon=2\varepsilon.
\end{equation} 
For convenience, we denote $G_\varepsilon^n:=G_\varepsilon(s+\rho_n)$, $n\in\mathbb{N}$. Due to $G(s)\in[1,3]$ for $s\in\mathbb{R}$, it follows form \eqref{eq2.37} that
\begin{equation}
	\label{eq2.38}	
	|G(s)+G_\varepsilon^n(s)|\leq G(s)+2\varepsilon+G(s)\leq6+2\varepsilon, \ s\in[nT_\varepsilon,(n+1)T_\varepsilon).
\end{equation}
By using $G(s)\in[1,3]$ and \eqref{eq2.37} again, we have
\begin{equation}
	\label{eq2.39}	
	G_\varepsilon^n(s/c)\geq G(s/c)-2\varepsilon\geq1-2\varepsilon\geq\frac{1}{2}, \ s\in[ncT_\varepsilon,(n+1)cT_\varepsilon).
\end{equation}

For $\Lambda(c)$ with $c>0$ in \eqref{eq2.33}, we can conclude from \eqref{eq2.37}, \eqref{eq2.38} and \eqref{eq2.39} that
\begin{equation}
	\label{eq2.40}	
	\begin{aligned}
		&\left|\Lambda(c)-\frac{1}{\pi}\sum_{n=0}^{\infty}\int_{ncT_\varepsilon}^{(n+1)cT_\varepsilon}\frac{\mathrm{d}s}{G^n_\varepsilon(\frac{s}{c})^2+s^2}\right|\\
		=&\left|\frac{1}{\pi}\left(\int_{0}^{cT_\varepsilon}+\cdots+\int_{ncT_\varepsilon}^{(n+1)cT_\varepsilon}+\cdots\right)\frac{\mathrm{d}s}{G(\frac{s}{c})^2+s^2}-\frac{1}{\pi}\sum_{n=0}^{\infty}\int_{ncT_\varepsilon}^{(n+1)cT_\varepsilon}\frac{\mathrm{d}s}{G^n_\varepsilon(\frac{s}{c})^2+s^2}\right|\\
		=&\frac{1}{\pi}\left|\sum_{n=0}^{\infty}\int_{ncT_\varepsilon}^{(n+1)cT_\varepsilon}\left(\frac{1}{G(\frac{s}{c})^2+s^2}-\frac{1}{G^n_\varepsilon(\frac{s}{c})^2+s^2}\right)\mathrm{d}s\right|\\
		=&\frac{1}{\pi}\left|\sum_{n=0}^{\infty}\int_{ncT_\varepsilon}^{(n+1)cT_\varepsilon}\frac{(G_\varepsilon^n(\frac{s}{c})-G(\frac{s}{c}))(G_\varepsilon^n(\frac{s}{c})+G(\frac{s}{c}))}{(G^n_\varepsilon(\frac{s}{c})^2+s^2)(G(\frac{s}{c})^2+s^2)}\mathrm{d}s\right|\\
		\leq&\frac{1}{\pi}\int_{0}^{\infty}\frac{2\varepsilon(6+2\varepsilon)}{((1-2\varepsilon)^2+s^2)(1+s^2)}\mathrm{d}s\\
		\leq&\frac{2\varepsilon(6+2\varepsilon)}{\pi}\int_{0}^{\infty}\frac{\mathrm{d}s}{(1-2\varepsilon)^2+s^2}\leq13\varepsilon.
	\end{aligned}
\end{equation}
From the periodicity of $G_\varepsilon$, we infer that 
\begin{equation}
	\label{eq2.41}	
	\begin{aligned}
		\sum_{n=0}^{\infty}\int_{ncT_\varepsilon}^{(n+1)cT_\varepsilon}\frac{\mathrm{d}s}{G^n_\varepsilon(\frac{s}{c})^2+s^2}=&\sum_{n=0}^{\infty}\int_{ncT_\varepsilon}^{(n+1)cT_\varepsilon}\frac{G^n_\varepsilon(\frac{s}{c})^2+s^2+(ncT_\varepsilon)^2-s^2}{(G^n_\varepsilon(\frac{s}{c})^2+s^2)(G^n_\varepsilon(\frac{s}{c})^2+(ncT_\varepsilon)^2)}\mathrm{d}s\\
		=&\sum_{n=0}^{\infty}\left(\int_{ncT_\varepsilon}^{(n+1)cT_\varepsilon}\frac{\mathrm{d}s}{G_\varepsilon(\frac{s}{c})^2+(ncT_\varepsilon)^2}+E_n\right)\\
		=&\sum_{n=0}^{\infty}\left(c\int_{0}^{T_\varepsilon}\frac{\mathrm{d}s}{G_\varepsilon(s)^2+(ncT_\varepsilon)^2}+E_n\right),
	\end{aligned}
\end{equation}
where
\begin{equation*}
	E_n:=\int_{ncT_\varepsilon}^{(n+1)cT_\varepsilon}\frac{(ncT_\varepsilon)^2-s^2}{(G^n_\varepsilon(\frac{s}{c})^2+s^2)(G^n_\varepsilon(\frac{s}{c})^2+(ncT_\varepsilon)^2)}\mathrm{d}s.
\end{equation*}
Note that $s\leq(n+1)cT_\varepsilon$. When $n\geq1$, we have $s\leq2ncT_\varepsilon$, which implies $ncT_\varepsilon\geq\frac{s}{2}$. It then follows from \eqref{eq2.39} that
\begin{equation*}
	\begin{aligned}
		|E_n|&\leq\int_{ncT_\varepsilon}^{(n+1)cT_\varepsilon}\frac{|(ncT_\varepsilon)^2-((n+1)cT_\varepsilon)^2|}{(\frac{1}{4}+s^2)(\frac{1}{4}+\frac{s^2}{4})}\mathrm{d}s\\
		&=\int_{ncT_\varepsilon}^{(n+1)cT_\varepsilon}cT_\varepsilon\frac{(2n+1)cT_\varepsilon}{\frac{1}{16}(1+4s^2)(1+s^2)}\mathrm{d}s\\
		&\leq48cT_\varepsilon\int_{ncT_\varepsilon}^{(n+1)cT_\varepsilon}\frac{s}{(1+s^2)^2}\mathrm{d}s, \ n\neq0.
	\end{aligned}
\end{equation*}
Moreover, when $n=0$, we get
\begin{equation*}
	|E_0|\leq\int_{0}^{cT_\varepsilon}\frac{s^2}{(\frac{1}{4}+s^2)\frac{1}{4}}\mathrm{d}s\leq16\int_{0}^{cT_\varepsilon}s^2\mathrm{d}s=\frac{16}{3}(cT_\varepsilon)^3.
\end{equation*}
Hence, letting $c\rightarrow0$ gives that
\begin{equation}
	\label{eq2.42}		
	\sum_{n=0}^{\infty}|E_n|\leq\frac{16}{3}(cT_\varepsilon)^3+48cT_\varepsilon\int_{cT_\varepsilon}^{\infty}\frac{s}{(1+s^2)^2}\mathrm{d}s\rightarrow0.
\end{equation}
For further calculation of \eqref{eq2.41}, we denote
\begin{equation*}
	F(t)=\frac{1}{T_\varepsilon}\int_{0}^{T_\varepsilon}\frac{\mathrm{d}s}{G_\varepsilon(s)^2+t^2}=\frac{1}{T_\varepsilon}\int_{0}^{T_\varepsilon}\frac{\mathrm{d}s}{G(s)^2+t^2}.
\end{equation*}
This means that the final summation of \eqref{eq2.41} can be written as
\begin{equation}
	\label{eq2.43}		
	\sum_{n=0}^{\infty}c\int_{0}^{T_\varepsilon}\frac{\mathrm{d}s}{G_\varepsilon(s)^2+(ncT_\varepsilon)^2}=\sum_{n=0}^{\infty}cT_\varepsilon F(ncT_\varepsilon).
\end{equation}
Treating the right hand side of \eqref{eq2.43} as a Riemann sum yields 
\begin{equation}
	\label{eq2.430}		
	\sum_{n=0}^{\infty}c\int_{0}^{T_\varepsilon}\frac{\mathrm{d}s}{G_\varepsilon(s)^2+(ncT_\varepsilon)^2}=\int_{0}^{\infty}F(t)\mathrm{d}t \ \text{as}\ c\rightarrow0.
\end{equation}
Substituting \eqref{eq2.42} and \eqref{eq2.430} into \eqref{eq2.41}, and combining with the Fubini's theorem, we see that 
\begin{equation}
	\label{eq2.44}	
	\begin{aligned}
		\lim_{c\rightarrow0}\frac{1}{\pi}\sum_{n=0}^{\infty}\int_{ncT_\varepsilon}^{(n+1)cT_\varepsilon}\frac{\mathrm{d}s}{G^n_\varepsilon(\frac{s}{c})^2+s^2}=&\frac{1}{\pi}\int_{0}^{\infty}F(t)\mathrm{d}t\\
		=&\frac{1}{\pi T_\varepsilon}\int_{0}^{T_\varepsilon}\int_{0}^{\infty}\frac{1}{G(s)^2+t^2}\mathrm{d}t\mathrm{d}s\\
		=&\frac{1}{\pi T_\varepsilon}\int_{0}^{T_\varepsilon}\frac{\pi}{2}\frac{1}{G(s)}\mathrm{d}s\\
		=&\frac{1}{2T_\varepsilon}\int_{0}^{T_\varepsilon}\frac{1}{G(s)}\mathrm{d}s.
	\end{aligned}
\end{equation}
Now, by allowing $\varepsilon$ to vary and taking the limit $\varepsilon \rightarrow 0$ in \eqref{eq2.40} and \eqref{eq2.44}, we deduce that $\Lambda(0^+)$ exists, and that $\lim_{c \rightarrow 0^+} \Lambda(c) = \gamma$.  This means that $\Gamma(0^-,\theta)$ exists, and $\lim_{c\rightarrow0^-}\Gamma(c,\theta)=\gamma(\theta)$ for all $\theta\in[0,2\pi)$.

\textit{Step 3}. The proofs of \textbf{(i)} and \textbf{(ii)}. From \eqref{eq22.1} and \eqref{eq2.250}, we see that \eqref{eq2.1} is equivalent to $\frac{\alpha}{\beta}-\frac{1}{2}=\Gamma(c,\theta)$ for $c\neq0$ and $\theta\in[0,2\pi)$. By \textit{Steps 1} and \textit{2}, it follows that 
\begin{equation*}
	|\Gamma(c,\theta)|<\frac{1}{2}, \ \frac{\partial\Gamma(c,\theta)}{\partial c}<0, \ \lim_{c\rightarrow-\infty}\Gamma(c,\theta)=\frac{1}{2} \ \text{and}\  \lim_{c\rightarrow0^-}\Gamma(c,\theta)=\gamma(\theta)
\end{equation*}
for all $\theta\in[0,2\pi)$ and $c\neq0$. The definition of $\Gamma(c,\theta)$ implies that it is odd with respect to $c\in\mathbb{R}$ for $\theta\in[0,2\pi)$. Therefore, we can conclude that \eqref{eq2.1} has a unique solution $c\neq0$ if and only if $|\frac{\alpha}{\beta}-\frac{1}{2}|>\gamma(\theta)$, and has no solution $c\neq0$ if and only if $|\frac{\alpha}{\beta}-\frac{1}{2}|\leq\gamma(\theta)$. This means that if $|\frac{\alpha}{\beta}-\frac{1}{2}|>\gamma(\theta)$, then \eqref{eq1.1} admits a bistable wave $U$ with $c\neq0$. If $|\frac{\alpha}{\beta}-\frac{1}{2}|\leq\gamma(\theta)$, then \eqref{eq1.1} admits no bistable waves $U$ with $c\neq0$. The proof of Theorem \ref{th1.2} is completed.

This section concludes with a more detailed analysis of the sign of $c$.  Note that $U'>0$ on $\mathbb{R}$. Combining \eqref{eq1.4} with $U(-\infty)=0$ and $U(\infty)=\beta$, we can conclude that 
\begin{equation*}
	c=\frac{\int_0^\beta f(U)\mathrm{d}U}{\int_{-\infty}^\infty(U'(\xi))^2\mathrm{d}\xi}.
\end{equation*}
This means that the sign of the nonzero $c$ is determined by $\int_0^\beta f(U)\mathrm{d}U$. From the definition of $f$, it follows that $c>0$ for $\frac{\alpha}{\beta}\in(0,\frac{1}{2})$, and $c<0$ for $\frac{\alpha}{\beta}\in(\frac{1}{2},1)$. In fact, we found a value $\gamma(\theta)\in(\frac{1}{6},\frac{1}{2})$ that can refine the interval of $\frac{\alpha}{\beta}$, thereby achieving a more precise characterization of $c$. Define
\begin{equation}
	\label{eq2.45}		
	\mathbb{A}_{\alpha,\beta}:=\left\{\theta\in[0,2\pi) \ |\  \gamma(\theta)<\left|\frac{1}{2}-\frac{\alpha}{\beta}\right|, \ \tan\theta\in\{\sqrt{3}r:r\in\mathbb{Q}\}\cup\{\pm\infty\} \right\}.
\end{equation}
Then we can conclude from Theorems \ref{th1.1} and \ref{th1.2} that the following corollary holds, which characterizes the critical conditions for propagation and pinning, as well as the propagation direction.

\begin{corollary}
	\label{co1}	
	Let $\mathbb{A}_{\alpha,\beta}$ be defined in \eqref{eq2.45}. 
	\begin{itemize}
		\item[$\mathrm{(i)}$] For  $\frac{\alpha}{\beta}\in(0,\frac{1}{3})$, if $\theta\in\mathbb{A}_{\alpha,\beta}$, then \eqref{eq1.1} admits a bistable wave with $c>0$; otherwise \eqref{eq1.1} admits no bistable waves with $c\neq0$. 
		\item[$\mathrm{(ii)}$] For  $\frac{\alpha}{\beta}\in(\frac{2}{3},1)$, if $\theta\in\mathbb{A}_{\alpha,\beta}$, then \eqref{eq1.1} admits a bistable wave with $c<0$; otherwise \eqref{eq1.1} admits no bistable waves with $c\neq0$.
		\item[$\mathrm{(iii)}$] For  $\frac{\alpha}{\beta}\in[\frac{1}{3},\frac{2}{3}]$, \eqref{eq1.1} admits no bistable waves with $c\neq0$.
	\end{itemize} 
\end{corollary}

\begin{figure}[ht]
\centering
{\includegraphics[width=3.4in,height=2.6in,clip]{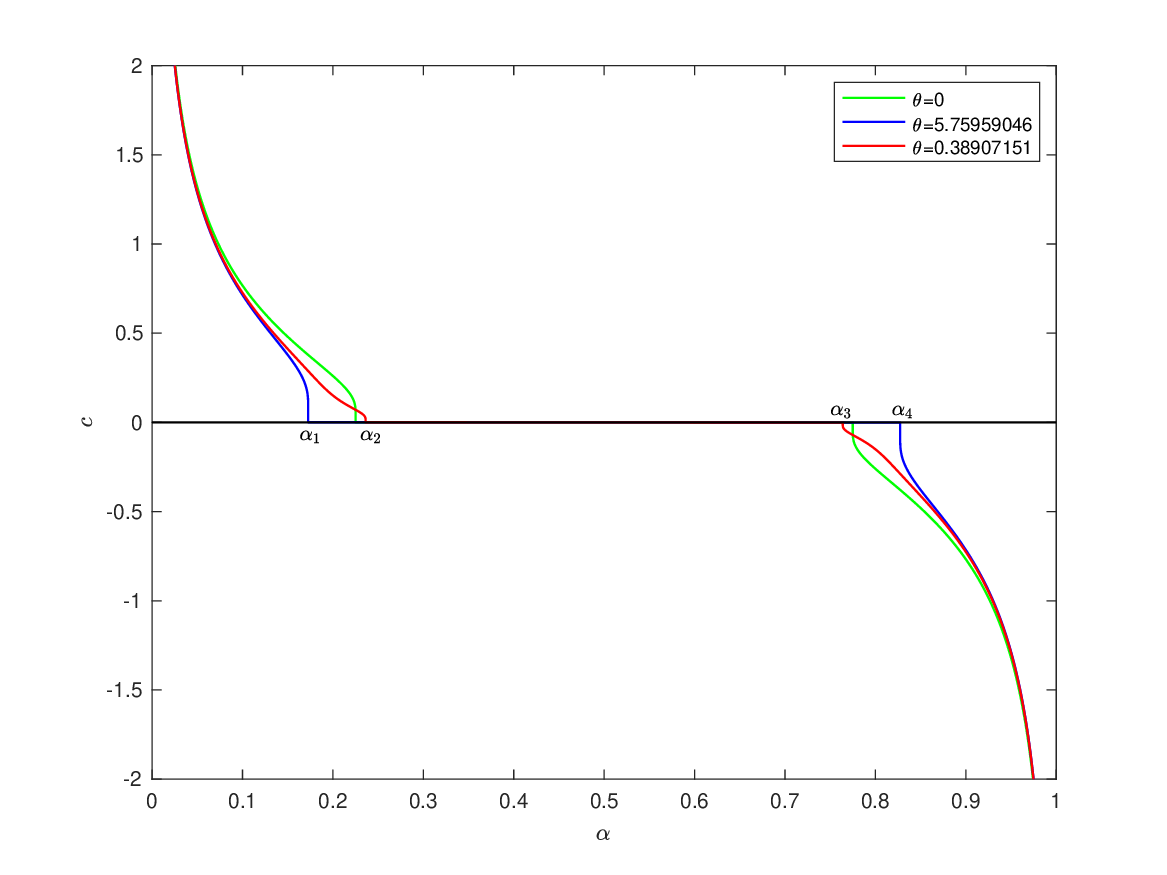} }	
\caption{The relationship between the wave speed $c$ and the parameter $\alpha$ in \eqref{eq2.1} for $\beta = 1$ and three distinct angles $\theta$. The intersections with the horizontal line $c=0$ correspond to the calculated thresholds $\alpha_k$, $k=1,2,3,4$.}\label{fig3}	
\end{figure}

To investigate the dependence of the wave speed	$c$ on the ratio $\alpha/\beta$, we define the auxiliary function $\Psi(\alpha, c,\theta):=\frac{\alpha}{\beta}-\frac{1}{2}-\Gamma(c,\theta)$ for $c\neq0$. The implicit function theorem implies that for fixed $\theta$ and $\beta$, there exists a differentiable function $c=c(\alpha)$ (when $c\neq0$) with derivative
\begin{equation*}	
c'(\alpha)=-\frac{\partial\Psi}{\partial\alpha}\frac{\partial c}{\partial\Psi}=\frac{1}{\beta}\frac{\partial c}{\partial\Gamma}<0,	
\end{equation*}		
indicating that the wave speed $c$ decreases monotonically with respect to $\alpha$.  To identify the specific propagation regimes, we define the following critical thresholds based on the extrema of $\gamma(\theta)$:
\begin{equation*}
	\alpha_1:=\frac{1}{2}-\gamma_{max},\ \alpha_2:=\frac{1}{2}-\gamma_{min},\
	\alpha_3:=\frac{1}{2}+\gamma_{min},\
	\alpha_4:=\frac{1}{2}+\gamma_{max},
\end{equation*}
where $\gamma_{max}:=\max_{\theta\in[0,2\pi)}\gamma(\theta)$ and $\gamma_{min}:=\min_{\theta\in[0,2\pi)}\gamma(\theta)$.
We numerically evaluate these extrema by discretizing the interval $[0,2\pi]$ into $400,000$ equidistant points and computing $\Gamma(-0.0006,\theta)$. Further calculation yields
\begin{equation*}
	\begin{aligned}	
		\gamma_{max}\approx\gamma(5.75959046)\approx0.32732328, \ 
		\gamma_{min}\approx\gamma(0.38907151)\approx0.26351684,\\
		\alpha_1\approx0.17267672, \ \alpha_2\approx0.23648316,\ \alpha_3\approx0.76351684, \ \alpha_4\approx0.82732328.
	\end{aligned}
\end{equation*}
\noindent \textbf{Numerical illustration of Corollary \ref{co1}.} To visualize the behavior of the wave speed relative to these thresholds, we set $\beta=1$ in \eqref{eq2.1} and plot the $c$-$\alpha$ relationship for selected propagation angles in Figure \ref{fig3}. The numerical results validate statements of Corollary~\ref{co1}: (i) $c>0$ for $\alpha\in(0,\alpha_1)$, the wave propagates successfully for all $\theta$; (ii) $c\geq0$ for $\alpha\in[\alpha_1,\alpha_2)$ and there exist angles where $c>0$ and others where $c=0$; (iii) $c=0$ for $\alpha\in[\alpha_2,\alpha_3]$ and the wave remains stationary for all propagation angles; (iv) $c\leq0$ for $\alpha\in(\alpha_3,\alpha_4]$ and the wave speed is $c<0$ for some $\theta$ and $c=0$ for others; and (v) $c<0$  for $\alpha\in(\alpha_4,1)$  and the wave propagates in the reverse direction for all $\theta$. Moreover, the monotonicity of the function $c=c(\alpha)$ on the intervals $(0,\alpha_1)$ and $(\alpha_4,1)$ is also illustrated.

\section{Angular dependence of the wave speed}\label{sec3}
	
\subsection{The relationship between $c$ and $\theta$}

\begin{figure}[ht!]
	\centering
	{\includegraphics[width=2.4in,height=2.4in,clip]{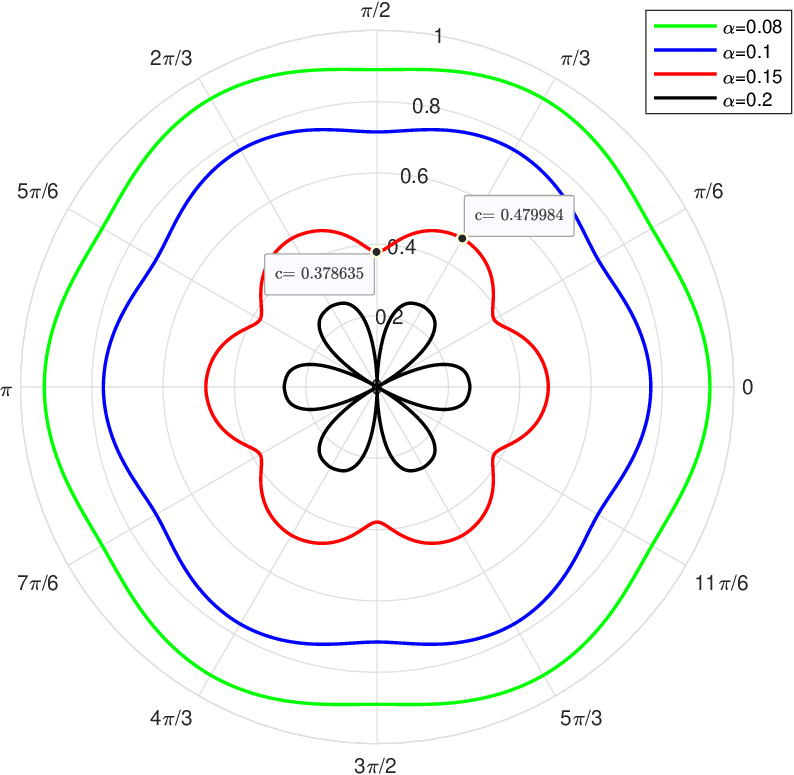} }	
	{\includegraphics[width=2.4in,height=2.4in,clip]{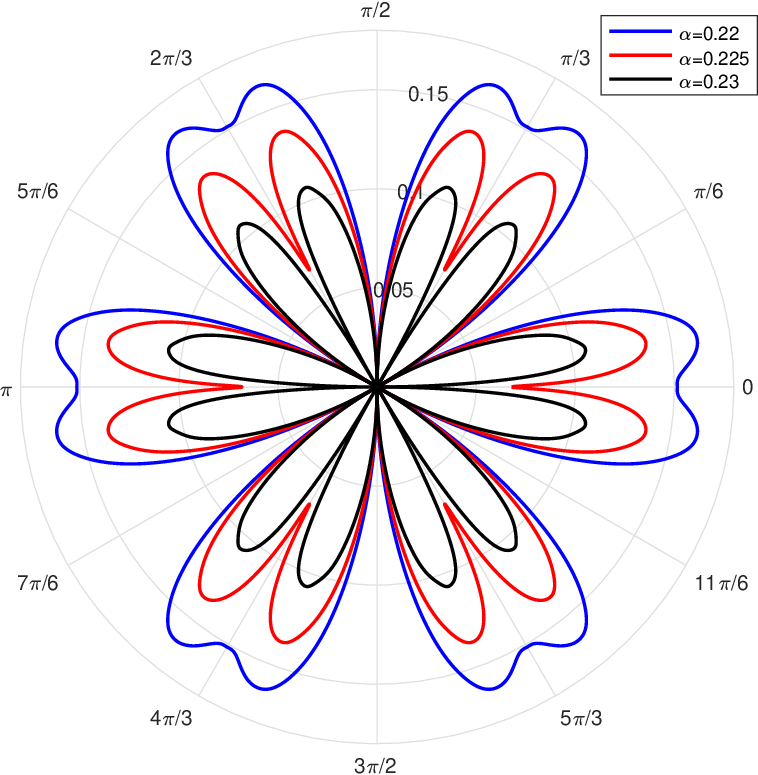} }	
	{\includegraphics[width=2.4in,height=2.4in,clip]{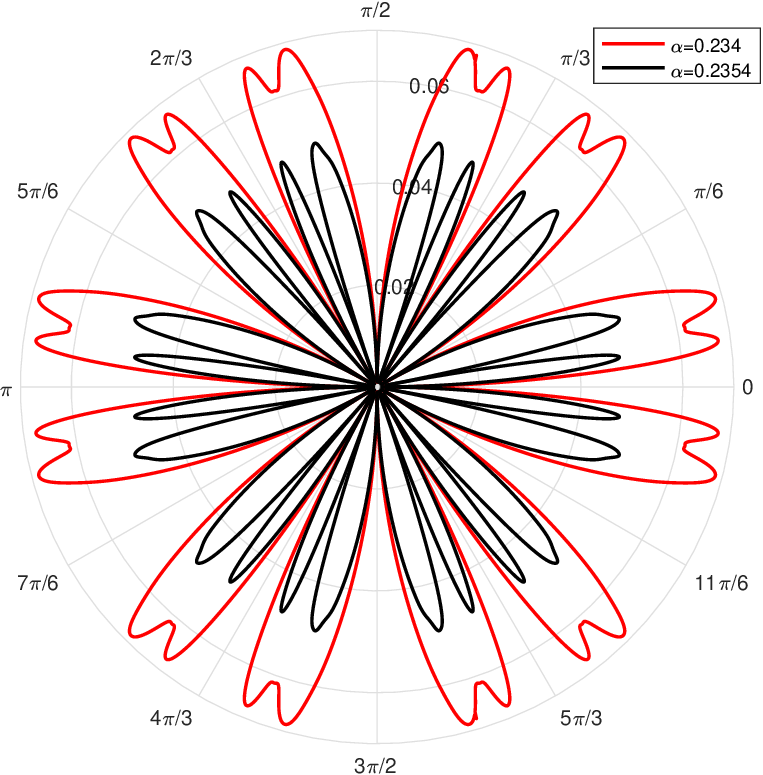} }	
	{\includegraphics[width=2.3in,height=2.4in,clip]{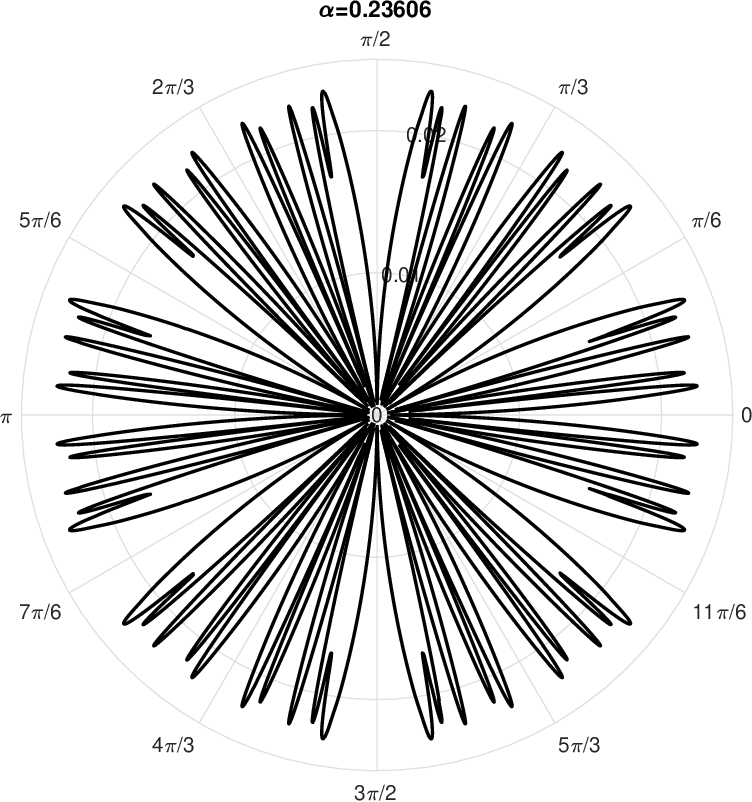} }
	\caption{The relationship between the absolute wave speed $|c|$ and angle $\theta$ in \eqref{eq2.1} for $\beta = 1$ and various values of $\alpha$.}\label{fig5}	
\end{figure}

The objective of this subsection is to investigate the dependence of the wave speed $c$ on the propagation angle $\theta$ and to provide numerical evidence validating the analytical findings in Corollary~\ref{co1}. Specifically, Figure~\ref{fig5} visualizes the variation of $c$ with respect to $\theta$ across various parameter regimes of $\alpha$, illustrating how this relationship changes across parameter regimes.	

According to Theorem \ref{th1.2}, the wave speed $c$ is characterized as the unique solution to the dispersion relation for a given $\theta$. By applying the implicit function theorem to the relation $\frac{\alpha}{\beta}=\frac{1}{2}-\Gamma(c,\theta)$, it can be shown that $c$ is a well-defined function of $\theta$. Furthermore, the structural symmetry of the function $G(s,\theta)$ implies that $c$ is $\frac{\pi}{3}$-periodic in $\theta$. Beyond angular symmetry, the wave speed exhibits a fundamental symmetry with respect to the parameter $\alpha$. Since $\Gamma(c,\theta)$ is an odd function of $c$, it follows from \eqref{eq2.1} that, for a fixed $\beta$, the wave speed $c$ is an odd function of the shift from the bistable balance point. Consequently, it suffices to restrict our investigation to $\alpha\in(0,\frac{\beta}{3})$ and focus on the behavior of non-negative wave speeds.	
	
We set $\beta=1$ and consider $\alpha\in (0, \alpha_2) \subset (0, \frac{1}{3})$. We sample $3,600$ equidistant points on the interval $[0, 2\pi]$ for the angular discretization. For this setup, we calculate the critical thresholds $\alpha_1 \approx 0.17665958$ and $\alpha_2 \approx 0.23610285$. By fixing representative values of $\alpha$ within these regimes, we plot the relationship between $c$ and $\theta$ in Figure~\ref{fig5}.

The numerical results in Figure~\ref{fig5} serve a twofold purpose. First, they verify the $\frac{\pi}{3}$-periodicity of the wave speed $c(\theta)$ resulting from the hexagonal lattice structure. Second, they illustrate how the qualitative behavior of $c(\theta)$ evolves as $\alpha$ increases. We partition the sampled values of $\alpha$ into two sets
\begin{equation*}
	\mathbb{E} := \{0.08, 0.1, 0.15\}, \quad \mathbb{F} := \{0.2, 0.22, 0.225, 0.23, 0.234, 0.2354, 0.23606\}.
\end{equation*}
For $\alpha \in \mathbb{E}$, as shown in the upper-left panel, the wave speed remains strictly positive $(c>0)$ for all $\theta \in [0, 2\pi)$. In this regime, $c$ is monotonically decreasing on the subinterval $[0, \frac{\pi}{6}]$ and monotonically increasing on $[\frac{\pi}{6}, \frac{\pi}{3}]$. This observation is consistent with the periodicity and monotonicity of the minimal wave speed established in previous studies of lattice dynamics (e.g., \cite[Theorem 2.3]{FLLW}). In contrast, for $\alpha \in \mathbb{F}$, the speed $c$ remains positive for certain angles but becomes almost zero for others, as depicted in the remaining three panels. As $\alpha$ increases toward the threshold $\alpha_2$, the angular dependence of $c$ becomes significantly more complex.

Finally, Figure~\ref{fig5} suggests that for certain ranges of $\alpha$, the bistable wave speed $c$ may exhibit piecewise monotonicity within specific angular subintervals. However, this feature is currently observed only numerically and lacks a rigorous theoretical justification. The increasing complexity for larger $\alpha$ may be related to the interplay between the $\theta$-periodicity and the almost periodicity of $G(s,\theta)$ in the variable $s$. A deeper analytical understanding of this directional complexity remains an important topic for future research.

\subsection{Numerical simulations of the wave profile and speed}

To visualize traveling wave solutions on the hexagonal lattice, this subsection presents numerical simulations of the wave profile and speed under varying positional and angular conditions. 

\begin{figure}[ht!]
	\centering
	{\includegraphics[width=3in,height=2.5in,clip]{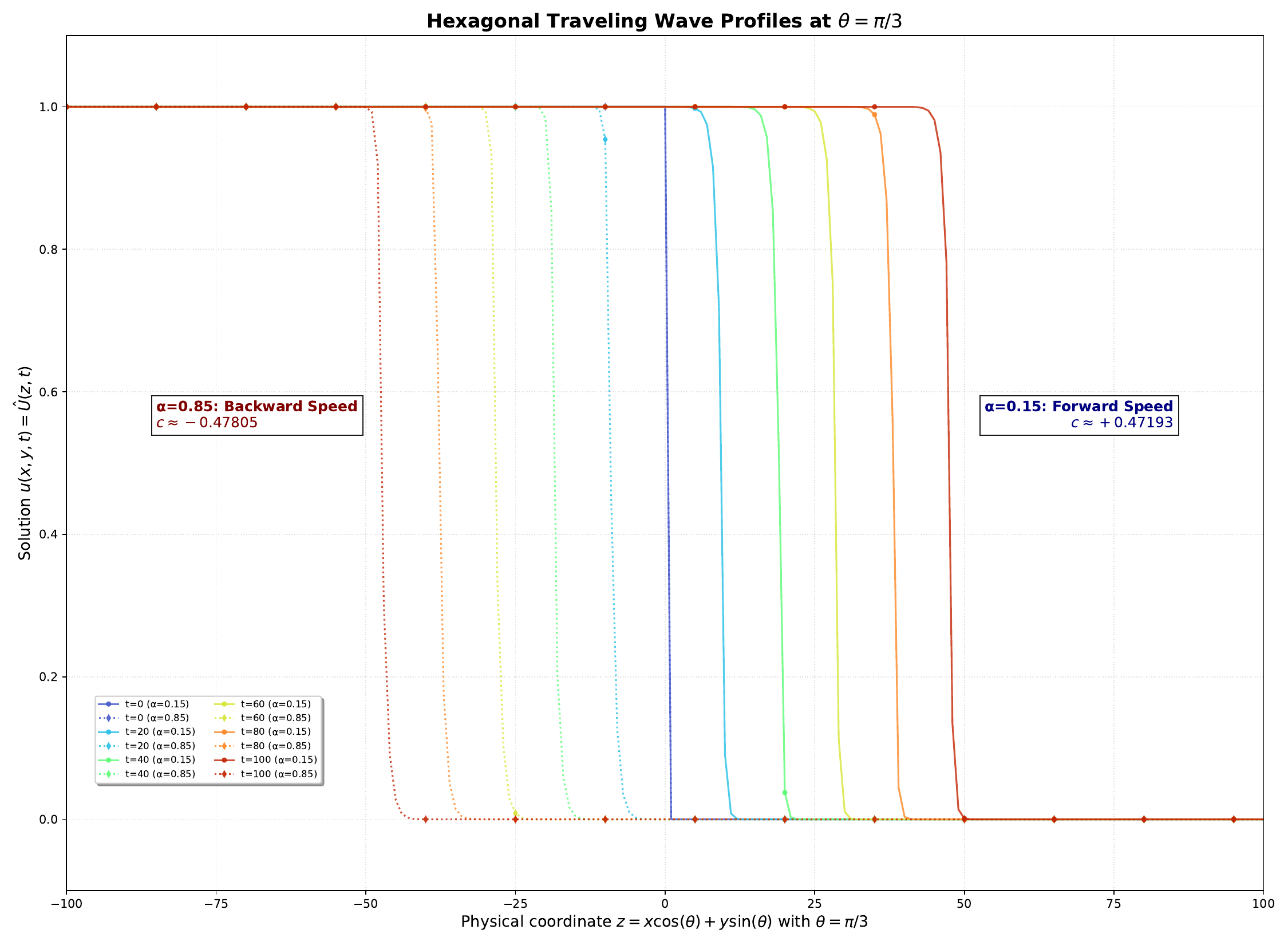} }	
	{\includegraphics[width=3in,height=2.5in,clip]{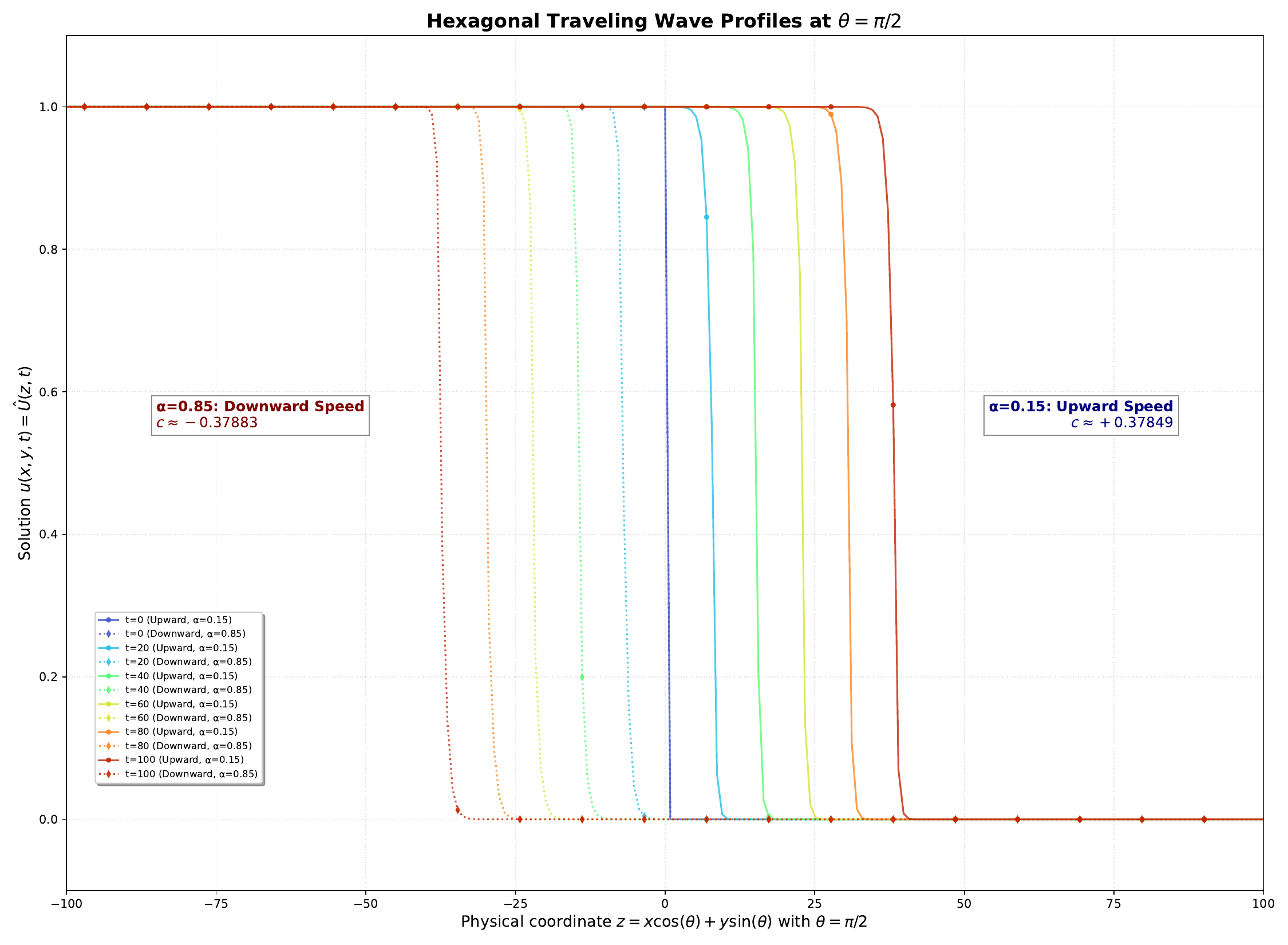} }	
	\caption{Wave profiles with nonzero speed at two different angles $\frac{\pi}{3}$ and $\frac{\pi}{2}$.}\label{fig6}	
\end{figure}

\begin{figure}[ht!]
	\centering
	{\includegraphics[width=5in,height=2.2in,clip]{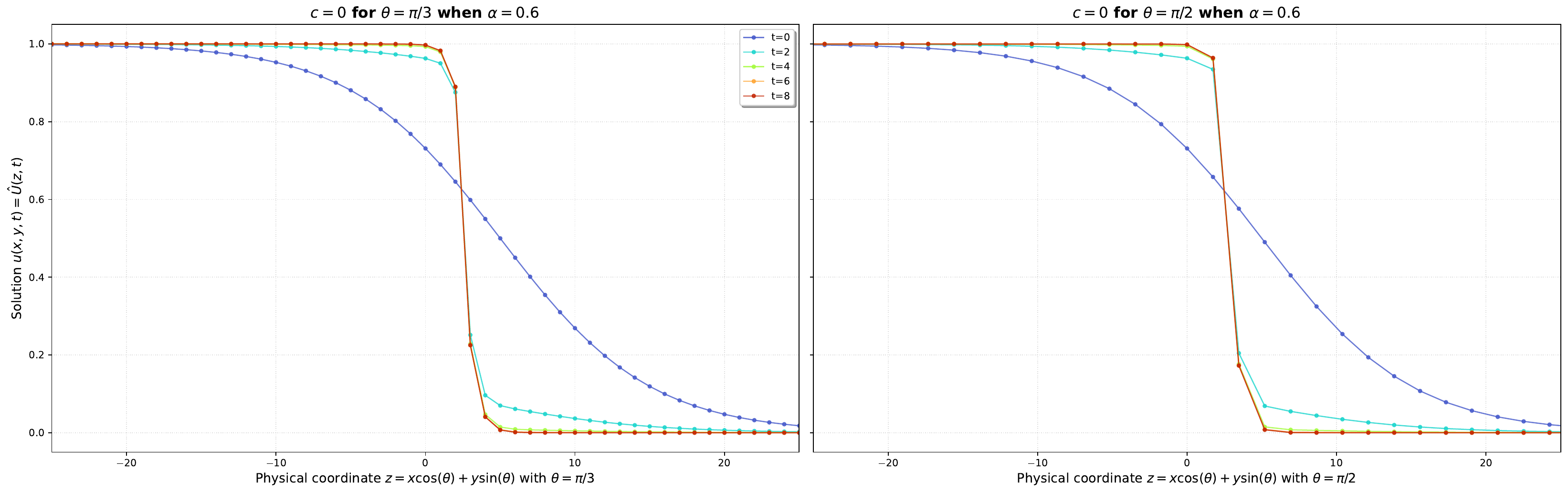} }	
	\caption{Pinning profiles at two different angles $\frac{\pi}{3}$ and $\frac{\pi}{2}$.}\label{fig7}	
\end{figure}

Recall that the form of traveling wave solutions $u(x,y,t)=U(x\cos\theta+y\sin\theta+ct)$ for $(x,y)\in\mathbb{H}$ and $\theta\in[0,2\pi)$. To provide numerical simulations of the wave profile $U$ propagating with speed $c$ over time $t$, we set 
\begin{equation*}
	U(x\cos\theta+y\sin\theta+ct):=(\hat{U},c), \ \ \hat{U}=\hat{U}(z,t), \ z=x\cos\theta+y\sin\theta.
\end{equation*}
We produce a planar wave on a two-dimensional domain by considering the initial distribution with a boundary cutting the domain connecting the stable and unstable states along a given direction. Furthermore, by setting $\beta=1$ and different $\alpha$, the plots of $(\hat{U},c)$ under different angles are presented in Figures \ref{fig6} and \ref{fig7}.

In Figure \ref{fig6}, as time $t$ increases, the initial distribution evolves to a traveling wave profile $\hat{U}$, with a fixed speed and a fixed shape. When $\alpha=0.85$, its speed is negative. When $\alpha=0.15$, its speed is positive. Moreover, the speed of the traveling wave depends on the choice of $\theta$. The numerical estimation of the speed matches well the theoretical values given in the upper-left panel of Figure \ref{fig5}, which are $c=0.479984$ for $\theta=\frac{\pi}{3}$, and $c=0.378635$ for $\theta=\frac{\pi}{2}$. From Figure \ref{fig7}, we consider $\alpha=0.6$, which leads to $c=0$ reflecting propagation failure, where solutions do not propagate and preserve their shape as time $t$ increases.

\section{Conclusion}\label{sec4}

The formulation of our mathematical model draws inspiration from propagation phenomena and hexagonal geometries in physics and cell biology. The model is an idealized reaction-diffusion equation posed on the hexagonal lattice. The main focus of this paper is to rigorously analyze the existence and speed of traveling wave solutions. More specifically, we identify an interval $|\frac{\alpha}{\beta}-\frac{1}{2}|>\gamma(\theta)$ in which propagation succeeds, and an interval $|\frac{\alpha}{\beta}-\frac{1}{2}|\leq\gamma(\theta)$ in which propagation fails. We also provide a detailed analysis of the continuity and discontinuity of $\gamma$ as a function of $\theta$. In particular, we examine how the sign of the wave speed depends on $\alpha$, $\beta$ and $\theta$. An interesting finding is that the wave speed varies with different directions, where propagation fails in specific directions, and these directions appear to be periodic in a complicated manner.

There are many directions where the current framework can be extended. In this work, we consider a specific nonlinearity to explicitly solve the one-dimensional traveling wave equation. How to develop the methods in this work to study the propagation dynamics of systems with more general bistable nonlinearity is an interesting extension. Another potential direction is to verify the theoretical results with specific physical or biological problems, for instance, the bistable optical chain \cite{AAC}. This may require us to adapt the current methods to nonsymmetrical coupling systems, which relates to challenging mathematical problems, such as the existence and speed of traveling wave solutions with spatially periodic nonlinearity \cite{DG}. Furthermore, it remains a problem to connect the propagation dynamics in discrete lattice systems to that in continuous models. For example, it was shown that the continuous equations with spatially heterogeneous nonlinearity can also generate propagation failure \cite{CER}. Moreover, obstacle problems in lattice differential equations are fundamental for understanding propagation dynamics and pattern formation in discrete media \cite{HHV}. The obstacle problem on the hexagonal lattice is a challenging direction.	

\section*{Acknowledgments}
Part of this work was completed while J. Wang was visiting The Hong Kong Polytechnic University, whose hospitality is gratefully acknowledged. Furthermore, the authors would like to thank the editor and the anonymous reviewers for their valuable comments and suggestions, which have greatly improved the quality of this paper.

\end{document}